%% file: main.tex
\documentclass[10pt,conference]{IEEEtran}

\usepackage{amsmath,amssymb,amsfonts}
\usepackage{cite}
\usepackage{graphicx}
\usepackage{textcomp}
\usepackage{multirow}
\usepackage{longtable}
\usepackage{lscape}
\usepackage{color, colortbl}
\usepackage{rotating}
\usepackage{wrapfig}
\usepackage{subcaption}
\usepackage{pdflscape}
\usepackage{hyperref}
\usepackage{array}
\usepackage{pifont}
\usepackage{balance}
\usepackage[most]{tcolorbox}
\usepackage{tikz}
\usepackage[normalem]{ulem}
\usepackage{url}
\usepackage{soul}
\usepackage{threeparttable}
\usepackage{xspace}
\usepackage{diagbox}
\usepackage{booktabs}
\usepackage{adjustbox}
\usepackage{svg}
\usepackage[ruled,vlined,lined,commentsnumbered]{algorithm2e}
\usepackage{amsmath,amsfonts}
\usepackage{csvsimple}
\usepackage{enumitem}
\usepackage{fancybox}
\usepackage{fontenc}
\usepackage{listings}
\usepackage{longtable}
\usepackage{lscape}
\usepackage{makecell}
\usepackage{marvosym}
\usepackage{moreverb}
\usepackage{multicol}
\usepackage{multirow}
\usepackage{pifont}% 
\usepackage{newtxmath}
\usepackage{rotating}
\usepackage{setspace}
\usepackage[most]{tcolorbox}
\usepackage{threeparttable}
\usepackage{tikz}
\usepackage[normalem]{ulem}
\usepackage{url}
\usepackage{soul}
\usepackage{wasysym}
\usepackage{indentfirst}
\usepackage{textcomp}
\usepackage{xcolor}
\usepackage{wrapfig}
\usepackage{pdflscape}
\usepackage{hyperref}
\usepackage{microtype}

\newcommand{\toolname}{{\sc AUGER}\xspace}
\newcommand{\toolnameb}{{\bf AUGER}\xspace}

\newcommand{\intuition}[1]{
\begin{tcolorbox}[colback=white,boxrule=1pt,top=0pt,bottom=0pt,left=1pt,right=2pt,top=2pt,bottom=2pt]
\em #1
\end{tcolorbox}
}

\begin{document}

\title{What You See Is What You Get: Attention-based Self-guided Automatic Unit Test Generation\vspace{-0.1cm}}

\author{
\IEEEauthorblockN{Xin Yin, Chao Ni\IEEEauthorrefmark{1}, Xiaodan Xu, Xiaohu Yang}
\IEEEauthorblockA{
\textit{The State Key Laboratory of Blockchain and Data Security, Zhejiang University, Hangzhou, China} \\
\{xyin, chaoni, xiaodanxu, yangxh\}@zju.edu.cn}
\vspace{-0.6cm}
}

\maketitle

\begingroup
\renewcommand\thefootnote{\IEEEauthorrefmark{1}}
\footnotetext{
This is the corresponding author\\Chao Ni is also with Hangzhou High-Tech Zone (Binjiang) Blockchain and Data Security Research Institute, Hangzhou, China}
\endgroup

\input{sections/abstract}

\noindent\vspace{-0.5cm}

\input{sections/introduction}
\label{sec:introduction}

\input{sections/motivation}
\label{sec:motivation}

\input{sections/approach}
\label{sec:approach}

\input{sections/experiment}
\label{sec:experiment}

\input{sections/results}
\label{sec:results}

\input{sections/threats}
\label{sec:threats}

\input{sections/related_work}
\label{sec:related_work}

\input{sections/conclusion}
\label{sec:conclusion}

\section*{Acknowledgements}{
This work was supported by the National Natural Science Foundation of China (Grant No.62202419), the Fundamental Research Funds for the Central Universities (No. 226-2022-00064),
Zhejiang Provincial Natural Science Foundation of China (No. LY24F020008),
the Ningbo Natural Science Foundation (No. 2022J184), 
and the State Street Zhejiang University Technology Center.
}

\bibliographystyle{IEEEtran}
\bibliography{main}

\end{document}

%% file: sections/abstract.tex
\begin{abstract}
Software defects heavily affect software's functionalities and may cause huge losses.
Recently, many AI-based approaches have been proposed to detect defects, which can be divided into two categories: software defect prediction and automatic unit test generation.
While these approaches have made great progress in software defect detection, they still have several limitations in practical application, including the low confidence of prediction models and the inefficiency of unit testing models.

To address these limitations, we propose a WYSIWYG (i.e., What You See Is What You Get) approach: 
\textbf{A}ttention-based Self-guided Automatic \textbf{U}nit Test \textbf{G}en\textbf{ER}ation (\toolname), which contains two stages: defect detection and error triggering.
In the former stage, \toolname first detects the proneness of defects.
Then, in the latter stage, it guides to generate unit tests for triggering such an error with the help of critical information obtained by the former stage. 
To evaluate the effectiveness of \toolname, we conduct a large-scale experiment by comparing with the state-of-the-art (SOTA) approaches on the widely used datasets (i.e., Bears, Bugs.jar, and Defects4J).
\toolname makes great improvements by 4.7\% to 35.3\% and 17.7\% to 40.4\% in terms of F1-score and Precision in defect detection, and can trigger 23 to 84 more errors than SOTAs in unit test generation.
Besides, we also conduct a further study to verify the generalization in practical usage by collecting a new dataset from real-world projects.
\end{abstract}

\begin{IEEEkeywords}
Defect, Unit Test Generation, Error-Triggering
\end{IEEEkeywords}

%% file: sections/introduction.tex
\vspace{-0.1cm}
\section{Introduction}

With the rapid advancement of industrial automation, code complexity and scale have increased, posing significant challenges in managing defects. 
Developers often miss potential defects unless explicit errors occur, reducing software quality. 
Recently, many approaches have been proposed to detect defects, which can be divided into two categories: software defect prediction and automatic unit test generation.
Despite the advancements in these approaches, they still face certain challenges and limitations.

Current defect detection approaches provide solely binary predictions~\cite{wardat2021deeplocalize, yamaguchi2014modeling, li2018vuldeepecker, li2021vuldeelocator, yang2024large, gao2018cobot, li2019deepfl, yin2024multitask}, indicating the presence of defects in code snippets or statements, but lack detailed explanations, making it challenging for developers to understand and trust the reliability of these predictions.  
To address this limitation, LineVul~\cite{fu2022linevul} uses attention scores for rationale, and some approaches incorporate the code's graph structure~\cite{zhou2019devign, hin2022linevd, li2021vulnerability, chakraborty2021deep, lou2021boosting, li2022fault}. 
However, these efforts only focus on explaining how models make specific decisions and neglect to provide detailed insights into the conditions that cause defects.
Meanwhile, Steenhoek et al.~\cite{steenhoek2023empirical} also show significant variability among different models under identical input conditions, eroding developers' trust in detection results.

Unit test generation approaches are crucial for ensuring software security and quality. 
Traditional unit test generation approaches prioritize achieving high code coverage.
However, research shows that high code coverage does not always trigger errors effectively~\cite{almasi2017industrial, shamshiri2015automatically}.
Recently, learning-based generation approaches~\cite{alagarsamy2023a3test, tufano2020unit, dinella2022toga} have made remarkable strides in progress. 
However, they focus on randomly generating a large number of test cases by training on the Method2Test dataset~\cite{tufano2022methods2test}, which contains numerous clean methods and non-error-triggering test cases, lacking effective information guidance. 
This leads to poor test efficiency, failing to efficiently trigger errors.

To address these limitations, we propose a WYSIWYG approach: \textbf{A}ttention-based Self-guided Automatic \textbf{U}nit Test \textbf{G}en\textbf{ER}ation (\toolname), which contains two stages: defect detection and error triggering.
In the first stage, \toolname detects defect proneness. 
In the latter stage, it guides the large language model (LLM) to generate unit tests for triggering such an error with the help of critical information obtained by the former stage.
As a result, \toolname will provide developers with defect detection results, convincing error-triggering unit tests, and corresponding test results. 
This automated approach instills confidence in developers regarding the detection results.

To investigate the effectiveness of \toolname, we first extract and filter methods from several widely used datasets~\cite{madeiral2019bears,saha2018bugs,just2014defects4j}, resulting in a total of 40,523 defective and non-defective methods.
Subsequently, we conduct comprehensive experiments to assess \toolname's performance in defect detection. 
The results indicate that \toolname can achieve an F1-score of 0.276, Precision of 0.198, and PR-AUC of 0.208 on Defects4J, which improves the baselines by 11.3\% to 35.3\%, 20.0\% to 40.4\%, and 24.6\% to 69.1\%, respectively.
Besides, we conduct extensive experiments to evaluate the effectiveness of \toolname in error triggering. 
The outcomes demonstrate that \toolname effectively triggers 35 and 84 errors in different scenarios, with a noticeably higher Precision than the baselines.
In our collection of real-world projects after March 2023, \toolname also achieves promising performance in error triggering.
Our main contributions are summarized as follows:

\textbf{A. WYSIWYG: Defect Detection with High Confidence:} \toolname provides defect detection results, convincing error-triggering unit tests, and corresponding test results, which helps to instill greater confidence in developers.

\textbf{B. Error Triggering with High Efficiency:}
\toolname guides the LLM to generate unit tests that trigger errors by leveraging critical information extracted from defective code, thereby narrowing down the search space.

\textbf{C. Extensive Evaluations:} We evaluate \toolname against current state-of-the-art approaches on the widely studied datasets~\cite{just2014defects4j, madeiral2019bears, saha2018bugs}.
To prevent data leakage, we also collect an additional real-world dataset for evaluation.
The replication package is publicly available at~\cite{replication}.

%% file: sections/motivation.tex
\section{Background and Motivation}

Defect detection and unit test generation are the main approaches to ensure software quality.
In this section, we aim to explore the challenges and limitations of existing defect detection approaches and unit test generation approaches.

\subsection{Limited confidence in defect detection models}

The limited confidence in defect detection models arises from two aspects: solely binary predictions and model inconsistency.

Current defect detection models often provide solely binary predictions, indicating whether a code snippet or statement contains defects or not~\cite{wardat2021deeplocalize, yamaguchi2014modeling, li2018vuldeepecker, li2021vuldeelocator, yang2024large, gao2018cobot, li2019deepfl,yin2024multitask,ni2024learning}.
These results lack detailed explanations, making it challenging for developers to understand and trust the reliability of these predictions.
LineVul~\cite{fu2022linevul} utilizes attention scores to explain the rationale behind the model's decision-making.  
Moreover, some approaches enhance the model's capabilities by incorporating the graph structure of code~\cite{zhou2019devign, hin2022linevd, li2021vulnerability, chakraborty2021deep, lou2021boosting, li2022fault}.
While these efforts have incorporated explanatory features, they often focus on why the model made a specific decision rather than providing detailed insights into the conditions that cause defects (e.g., the inputs and outputs that can expose defects).

Meanwhile, the inconsistency of the defect detection models also undermines developers' trust.
Steenhoek et al.~\cite{steenhoek2023empirical} reveal significant variability among different models, as they may produce entirely divergent results under identical input conditions, highlighting a lack of consistency. 
Consistency indicates that for the same input (e.g., one function), all models in the study give the same prediction (e.g., for a specific function, all methods predict it as defective or all agree that it is non-defective). Consistency can bring trust in prediction when developers make decisions.
To assess the consistency of different models in defect detection, we also conducted an empirical study, as shown in Table~\ref{tab:motivation}.
Specifically, we fine-tuned two learning-based detection models (i.e., LineVul~\cite{fu2022linevul} and SVulD~\cite{ni2023distinguishing}) and two pre-trained models (i.e., UniXcoder~\cite{guo2022unixcoder} and CodeBERT~\cite{feng2020codebert}) respectively on three widely used defect detection datasets~\cite{madeiral2019bears, saha2018bugs, just2014defects4j}, and compute the consistency in different models.
We can see that the consistency of all models is only 51.0\% to 58.5\%.
Such low consistency will erode developers' trust in the defect detection models, which hinders their practical application.

\begin{table}[htbp]
\centering
\caption{Consistency in different defect detection models}
\vspace{-0.3cm}
\begin{center}
\resizebox{.8\linewidth}{!}
{
    \begin{tabular}{lcc}
    \toprule
    \textbf{Models} & \textbf{Bears+Bugs.jar} & \textbf{Defects4J} \\
    \midrule
        Learning-based models & 71.7\% & 65.0\% \\
        Pre-trained models & 81.3\% & 79.5\%  \\
        All models & 58.5\% & 51.0\% \\
    \bottomrule
    \end{tabular}
}
\vspace{-1cm}
\label{tab:motivation}
\end{center}
\end{table}

\textcolor{black}
{
\subsection{Limited efficiency in unit test generation approaches}
}

Unit test generation approaches have gained widespread attention and application due to their ability to directly identify defects in software.
Existing unit test generation approaches can be categorized into two types: traditional generation approaches and learning-based generation approaches. 
However, both of them suffer from inefficiency issues.

Traditional generation approaches~\cite{fraser2011evosuite, pacheco2007randoop} focus on the code coverage metric. 
Researches show that traditional generation approaches are very effective at achieving high coverage~\cite{aleti2017analysing, oliveira2018mapping, panichella2015reformulating, panichella2017automated}, even covering more code than manually written test cases.
However, previous studies indicate that high code coverage does not always result in effective error triggering~\cite{almasi2017industrial, shamshiri2015automatically}.

Meanwhile, learning-based generation approaches~\cite{alagarsamy2023a3test, tufano2020unit, dinella2022toga} have been proposed and have made significant progress.
These approaches are trained on specific test generation datasets (e.g., Methods2Test~\cite{tufano2022methods2test}), enabling them to generate test cases that meet specific testing objectives.
They focus on randomly generating a large number of test cases, lacking effective information guidance. 
This leads to poor test efficiency, failing to efficiently trigger errors.
Moreover, a dataset comprising high-quality pairings of defective methods with error-triggering test cases is essential for training a model that efficiently generates error-triggering test cases.
However, existing approaches are typically trained on the Methods2Test~\cite{tufano2022methods2test} dataset, which contains numerous clean methods and test cases incapable of triggering errors. 
Consequently, a significant portion of the test cases generated by these approaches fail to trigger errors, resulting in limited efficiency in error triggering.

\vspace{0.2cm}
\intuition{{\bf Intuition.}
Providing corresponding error-triggering unit tests along with test results will help to instill greater confidence in defect detection results. 
Simultaneously, defect detection information can be leveraged to guide the generation of unit tests, reducing the model's search space and enhancing the efficiency of unit test generation.
}

%% file: sections/approach.tex
% \vspace{-0.2cm}
\section{Approach}

\begin{figure*}[htbp]
    \vspace{-0.3cm}
    \centering
    \includegraphics[width=.83\linewidth]{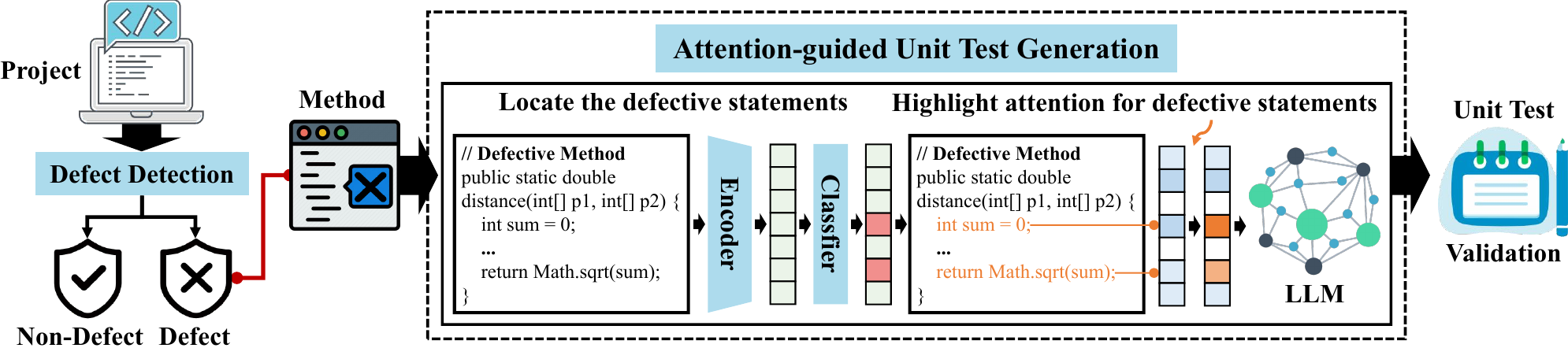}
    \caption{Overview of \toolname}
    \vspace{-0.5cm}
    \label{fig:overview}
\end{figure*}

We propose a WYSIWYG approach: {\textbf{A}ttention-based Self-guided Automatic \textbf{U}nit Test \textbf{G}en\textbf{ER}ation (\toolname)}, which contains two stages: defect detection and error triggering.
In the former stage, \toolname first detects the proneness of defects.
In the latter stage, it guides the LLM to generate unit tests for triggering such an error with the help of critical information present in defective code.
As a result, \toolname will provide developers with defect detection results, convincing error-triggering unit tests, and corresponding test results. 
Although \toolname is general, in this paper, we adpot recent DeepSeek Coder~\cite{deepseek-coder} and CodeLlama~\cite{roziere2023code} as the backend LLMs, which can be easily replaced with various state-of-the-art models (e.g., CodeT5+~\cite{wang2021codet5} and StarCoder~\cite{li2023starcoder}). 
\toolname includes defect detection, attention-guided unit test generation, and unit test validation.
Fig.~\ref{fig:overview} provides an overview of our approach:

\begin{itemize}[leftmargin=*]
\item \textbf{Defect Detection}. 
We first introduce a Java class file and extract all methods in the class file using the JavaParser~\cite{JavaParser} tool.
We encode the methods into token representations and input them into \toolname to detect whether there are defects.
\item \textbf{Attention-guided Unit Test Generation}. 
We conduct further analysis on the detected defective methods to identify the defective statements.
Then, we guide the LLM to focus on the defective statements in order to generate unit tests that trigger the errors. 
\item \textbf{Unit Test Validation}.
We describe how \toolname injects a unit test into an existing suite (i.e., test class) and elaborate on how \toolname adds the required dependencies to successfully execute the injected unit test.
\end{itemize}

\subsection{Defect Detection}

To improve robustness and performance, we adopt adversarial learning and contrastive learning framework with the pre-trained model, UniXcoder~\cite{guo2022unixcoder}. 
There are three important components of \toolname: (1) an encoder for embedding methods' semantics, (2) an attack strategy for generating adversarial samples, and (3) a learning strategy for discriminating differences between normal and adversarial samples.

\subsubsection{Code Encoder}
UniXcoder, proposed by Guo et al.~\cite{guo2022unixcoder}, is a unified pre-trained model incorporating semantic and syntax information from both code comment and AST, and we adopt it as the code encoder in \toolname to embed the code features at the method level.
UniXcoder transforms the input method to a 768-dimensional embedding $E_{M}$~\cite{liu2023enhancing,liu2024beyond}.

\subsubsection{Adversarial Learning}
In adversarial learning, the goal is to enhance the robustness of the model against adversarial attacks. 
We employ FGM~\cite{miyato2016adversarial} to conduct adversarial learning on UniXcoder, producing the encoded samples after the attack, denoted as $E_{A}$.
The process involves: 
(1) Perturbations are applied to the original input data to generate adversarial samples. These perturbations are crafted to challenge the model’s ability to correctly classify or handle the input.
(2) Training with Adversarial Samples: The model is trained on both the original and adversarial samples.
This training helps the model learn to recognize adversarial inputs.

\subsubsection{Contrastive Learning} 
Contrastive learning~\cite{ni2023distinguishing, wu2021r, gao2021simcse} is employed to enhance the model’s resistance to perturbations by aligning normal and adversarial samples. 
The training objective is to fine-tune the network such that the encoded representations $E_{M}$ (i.e., normal samples) and $E_{A}$ (i.e., adversarial samples) are as close as possible.
The objective can be described as $max(||E_M - E_A||, 0)$, which encourages the network to reduce the distance between these representations.

We employ the KL-divergence loss used in R-Drop~\cite{wu2021r} to quantify the differences between normal and attacked samples to minimize the distance between $E_{M}$ and $E_{A}$.
During the fine-tuning phase, we use the cross-entropy loss to guide the optimization process of \toolname by comparing the difference between the prediction probability of the model and the label.
The final loss of defect detection consists of both classification cross-entropy loss and KL-divergence loss, which can be described by the following equation:

\vspace{-0.2cm}
\begin{equation}\label{eq1}
\mathcal{L}_{ce} = -\sum_{i} y_i \log(\hat{y}_i)
\end{equation}
\begin{equation}\label{eq2}
\mathcal{L} = \mathcal{L}_{ce} + \beta \cdot \mathcal{L}_{kl}
\end{equation}

\vspace{-0.3cm}
\subsection{Attention-guided Unit Test Generation}
LLM is pre-trained using millions of code snippets from open-source projects, showing dominantly superior reasoning capabilities over existing AI models in downstream tasks~\cite{yin2024enhancing, bang2023multitask, yin2024rectifier, ouyang2022training, yin2024thinkrepair}.
In this section, we aim to stimulate the powerful capabilities of LLM to efficiently generate error-triggering unit tests.
To achieve this, we need to address four tasks: \textbf{(1) Defect Location}, \textbf{(2) Prompt Preparation}, \textbf{(3) Attention Profiling}, and \textbf{(4) Attention Inference}.

\subsubsection{Task 1: Defect Location}

Similar to the defect detection process, we also utilize UniXcoder as the foundational model for defect location, adopting it to embed code features at the statement level.
More precisely, given the source code of a defective method, \toolname first splits the method into individual statements.
Then, \toolname transforms the input method to $n\times768$-dimensional vectors at the statement level, where $n$ indicates the number of statements. 
Finally, after passing the final representation through a fully connected layer and the $softmax$ layer, \toolname outputs defective statements.
The cross-entropy loss function is used to train the process.

\begin{figure}[htbp]
    \centering
    \vspace{-0.2cm}
    \includegraphics[width=.65\linewidth]{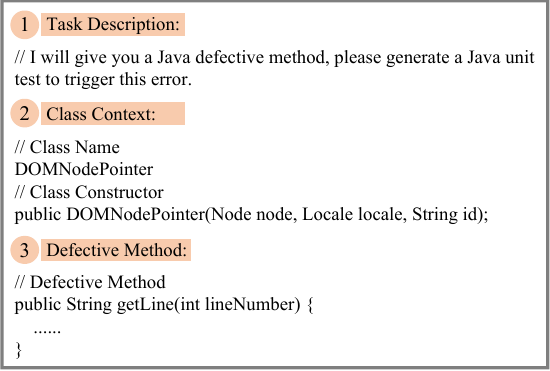}
    \caption{An example of prompt for LLM}
    \vspace{-0.2cm}
    \label{fig:prompt}
\end{figure}

\subsubsection{Task 2: Prompt Preparation}
\toolname is generalizable and can be extended to other programming languages by 
modifying the language-specific information in the prompt (e.g., \textit{``Java''} to \textit{``Python''} and \textit{``//''} to \textit{``\#''}).
The prompt involves three important components as illustrated in Fig.~\ref{fig:prompt}:

\begin{itemize}[leftmargin=*]
\item \textbf{Task Description} (marked as \ding{172}).
LLM is provided with the description constructed as \textit{``// I will give you a Java defective method, please generate a Java unit test to trigger this error''}. 
\item \textbf{Class Context} (marked as \ding{173}).
The class name and class constructor provide key information about the structure and initialization process of the class.
\item \textbf{Defective Method} (marked as \ding{174}). 
We provide the defective method in the method-level error-triggering scenario. 
We also prefix the defective method with \textit{``// Defective Method''} to directly indicate LLM about the context of the method.
\end{itemize}

Due to the vast search space of LLMs, these models randomly generate entirely different unit tests for focal methods.
To efficiently obtain high-quality outputs within a reasonable time frame, we have modified the attention mechanism of the LLM.
This modification is designed to guide the model's attention specifically toward the statements where defects are located.
The specific process (Algorithm~\ref{alg:pasta}) consists of two components: \textbf{(1) Attention Profiling}, which selects the effective attention heads for modifying, and \textbf{(2) Attention Inference}, which emphasizes the defect location information of the defective methods during inference.

\IncMargin{-0.5em}
\begin{algorithm} 
\small
    \textbf{Attention Profiling (Section \ref{sec:profiling})}
    
    \textbf{Input: }Profiling set $\mathcal{D}$, coefficient $\alpha$, attention layer number $L$, attention head number $H$, hyperparameter $k$;
    
    \For{$l\leftarrow 1$ \KwTo $L$, $h\leftarrow 1$ \KwTo $H$} {
        \emph{1: Modify the attention head $(l, h)$ by Equation~\ref{eq:eq4};}
        
        \emph{2: Evaluate the unit test generation performance on $\mathcal{D}$;}
        
        \emph{3: Collect the top $k$ heads $\mathcal{H}$ with the validation results;}
    }
    
    {\textbf{Output: }The attention head set $\mathcal{H}$;}

    \vspace{0.1cm} 
    
    \hrule

    \vspace{0.1cm} 
    
    \textbf{Attention Inference (Section \ref{sec:inference})}
    
    \textbf{Input: }Prompts $\mathcal{P}$, defective statements $\mathcal{S}$, coefficient $\alpha$;
    
    \For{head $(l, h)$ in $\mathcal{H}$} {
        \emph{1: Modify the attention head $(l, h)$ by Equation~\ref{eq:eq4};}
        
        \emph{2: Generate a large number of candidate unit tests;}
    }
    
    \textbf{Output: }Unit tests $\mathcal{T}$;
    \caption{Attention Profiling and Attention Inference}
    \label{alg:pasta}
\end{algorithm}

\subsubsection{Task 3: Attention Profiling}
\label{sec:profiling}

LLM typically has multiple attention layers and multiple attention heads (e.g., DeepSeek Coder 6.7B has 32 attention layers, each with 32 attention heads).
Attention heads~\cite{vaswani2017attention} are components of the multi-head attention mechanism in Transformer models.
Each attention head operates independently, enabling the model to focus on different parts of the input sequence simultaneously. 
However, there is currently no consensus on the specific roles that these attention heads play.
Zeng et al.~\cite{zeng2022extensive} demonstrated that the first attention layer provides insight into which tokens the model focuses on. 
On the other hand, Wan et al.~\cite{wan2022they} showed that deeper attention layers excel in capturing long-distance dependencies and program structure.
It is important to specify the correct attention heads, given that different heads serve distinctive roles in encoding semantic/syntactic information.
To this end, we propose an attention profiling algorithm to identify the effective attention heads for inference. 
Specifically, we sub-sample profiling set $\mathcal{D}$ (100 samples) from the Defects4J dataset (cf. Section~\ref{sec:dataset}).
After that, we need to define how to modify the LLM's attention so that it focuses on the specified statement. 
\toolname emphasizes the defective statements of the input method by down-weighting the attention scores of tokens that are not in the defective statements.
Specifically, given the tokens of highlighted statements as $\mathcal{S}$,
\toolname emphasizes these tokens by an attention projection $\mathcal{W}$:

\vspace{-0.4cm}
\begin{equation}
\small
\label{eq:eq4}
\boldsymbol{H}^{(l, h)}=\mathcal{W}\left(\boldsymbol{A}^{(l, h)}\right) \boldsymbol{V}, \text{where } [\mathcal{W}(\boldsymbol{A})]_{t}= \begin{cases} \boldsymbol{A}_{t} / C & \text{if} \ t \in \mathcal{S} \\ \alpha\boldsymbol{A}_{t} / C & \text {else}\end{cases}
\end{equation}
\vspace{-0.3cm}

where $0 \leq \alpha<1$ is a scaling coefficient and $\boldsymbol{A}^{(l, h)}$ denotes the attention scores at the head $h$ of the $l$-th layer.
The term $C=\sum_{t \in \mathcal{S}} \boldsymbol{A}_{t}+\sum_{t \notin \mathcal{S}} \alpha \boldsymbol{A}_{t}$ normalizes the scores so that they sum up to one. 
Attention modifying is conducted during the inference process and does not require any training.

\begin{figure}[htbp]
    \centering
    \vspace{-0.3cm}
    \includegraphics[width=.85\linewidth]{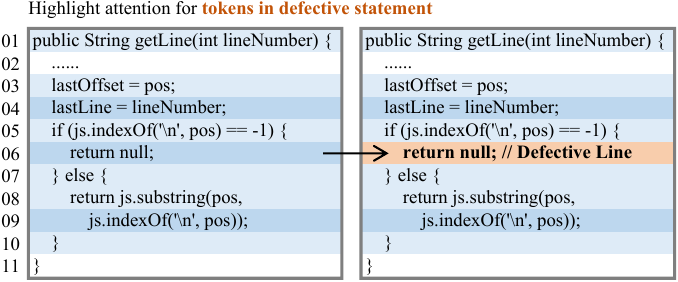}
    \caption{Highlight attention for tokens in defective statement}
    \vspace{-0.3cm}
    \label{fig:attention}
\end{figure}

Equation~\ref{eq:eq4} modifies the model attention by scaling down the scores of tokens that are not in defective statements.
When the coefficient $\alpha$ is set very small, defective statements are highlighted given their increased attention scores after re-normalization.
As shown in Fig.~\ref{fig:attention}, \toolname recognizes that line 6 is a defective statement.
Consequently, \toolname first marks the defective line in the prompt, and then recalculates the attention weights within the attention heads, directing the LLM's focus towards the tokens associated with the statement.

After that, we assess the performance of modifying each individual attention head $(l, h)$, where $1\leq l \leq L$ and $1 \leq h \leq H$, on a designated subset $\mathcal{D}$. 
We rank all the heads based on their unit test generation performance, specifically, by evaluating how many errors can be triggered on $\mathcal{D}$.
Subsequently, we define the attention head set $\mathcal{H}$ for inference as the top $k$ performing heads.

Unlike fine-tuning, attention profiling doesn't modify any model weights, so it demands similar computational resources as inference. 
The resulting head set $\mathcal{H}$ serves as a model-level profile. 
Once determined, we can use attention inference on $\mathcal{H}$ for both existing and unseen datasets, enhancing model comprehension and boosting performance.

\subsubsection{Task 4: Attention Inference}
\label{sec:inference}
During the inference process, \toolname modifies the attention head $(l, h)$ in the attention head set $\mathcal{H}$ using Equation~\ref{eq:eq4}. 
The input to this process includes the prompt (i.e., task description, class context, and defective method), defective statement, and the coefficient $\alpha$, while the output consists of a large number of candidate unit tests.

\subsection{Unit Test Validation}
Our approach is automated and requires no manual intervention. 
Therefore, to validate whether the unit tests generated by LLM can accurately trigger identified errors, we need to automatically inject the unit tests into the corresponding test classes. 
Additionally, we must add the required dependencies to execute and obtain the test results.

\subsubsection{Inject unit test into test class}
We use token similarity to find the test classes that are most similar to the generated unit tests and inject them.
The intuition is that, if a unit test belongs to a test class, the unit test likely uses similar methods and classes, and it shares similar tokens to other tests from that test class. 
Formally, we assign a matching score for each test class based on equation: $\operatorname{sim}_{c_i}=\left|T_t \cap T_{c_i}\right| /\left|T_t\right|$, where $T_t$ and $T_{c_i}$ are the set of tokens in the generated unit test and the $i$-th test class, respectively.

\subsubsection{Add the required dependencies} 
First, \toolname parses the generated unit test and identifies variable types and the referenced class names, constructors, and exceptions.
\toolname then endeavors to locate public classes matching the identified type name for unimported dependencies. 
If precisely one such file exists, \toolname derives the classpath to the identified class and adds an import statement accordingly.
However, there may be scenarios where no matching classes are found or multiple matches occur. 
In such cases, \toolname scans the project for \textit{import} statements ending with the target class name and selects the most prevalent import statement across all files.

After injecting the unit test into the test class and adding the required dependencies, \toolname executes the test to check whether it triggers the identified errors.

%% file: sections/experiment.tex
\section{Experiment}
In this section, we first present the studied datasets, and then introduce the baseline approaches.
Following that, we describe the performance metrics as well as the experimental setting.

\subsection{Datasets}
\label{sec:dataset}

\subsubsection{Defect Detection}
In order to ensure the thoroughness and validity of our research findings regarding defect detection, we have leveraged three widely used Java defect datasets: the \textbf{Bears} dataset~\cite{madeiral2019bears}, the \textbf{Bugs.jar} dataset~\cite{saha2018bugs}, and the \textbf{Defects4J} dataset~\cite{just2014defects4j}.

Since \toolname focuses on method level, we perform two filtering steps on the original datasets to obtain valid methods, and the filtering results of each dataset are displayed in Table~\ref{tab:dataset}.

\textbf{Step-1:} Each commit is considered as a mini-version of a project.
We use the commit IDs to request commit histories of the projects, and for each commit, we extract the code changes between before and after fixing a defect.
Finally, we use the code change information to obtain the defective and fixed version of a method. 
Thus we collect the following information for a project: defective methods with their fixes and other clean methods.
In this step, we obtain the Bears dataset, consisting of 2,009 methods, the Bugs.jar dataset, consisting of 40,880 methods, and the Defects4J dataset, containing 31,423 methods.

\textbf{Step-2:} To clean and normalize the dataset, we start by removing duplicate methods. 
The three datasets are derived from various versions of projects (e.g., Defects4J extracted from 17 real-world Java projects), leading to a substantial number of duplicates in methods extracted from different commits during step-1.
In this step, we finally obtain the Bears dataset, which comprises 1,769 methods, the Bugs.jar dataset, which comprises 20,948 methods, and the Defects4J dataset, which comprises 17,806 methods.

\subsubsection{Unit Test Generation}
Defects4J includes utilities for generating and evaluating test suites on the programs to determine if generated tests pass on the fixed versions and catch defects on the defective versions.
In contrast, Bears and Bugs.jar do not include readily executable test suites.
Therefore, we evaluate real-world error triggering on the Defects4J dataset.

\begin{table}[htbp]
\centering
\caption{The statistic of studied datasets}
\vspace{-0.2cm}
\begin{center}
\resizebox{.8\linewidth}{!}
{
\begin{tabular}{lrrrr}
\toprule
\textbf{Datasets} & \textbf{\# Defective} & \textbf{\# Clean} & \textbf{\# Total} & \textbf{\% Ratio} \\
\midrule
Bears & 132 & 1,637 & 1,769 & 7.46\% \\
Bugs.jar & 1,953 & 18,995 & 20,948 & 9.32\% \\
Defects4J & 1,130 & 16,676  & 17,806 & 6.34\% \\
\bottomrule
\end{tabular}
}
\vspace{-0.5cm}
\label{tab:dataset}
\end{center}
\end{table}

\subsection{Baselines}

\subsubsection{Defect Detection}
To comprehensively compare the performance of existing work, we consider two learning-based detection approaches and two pre-trained models.
The former group contains two approaches (i.e., LineVul~\cite{fu2022linevul} and SVulD~\cite{ni2023distinguishing}), which simply treat the source code as a sequence of tokens to represent its semantics.
The latter group contains two models (i.e., CodeBERT~\cite{feng2020codebert} and UniXcoder~\cite{guo2022unixcoder}), which are pre-trained models for programming languages.

\subsubsection{Unit Test Generation}
We consider the following baselines in this evaluation:

\textbf{Traditional Approach.}
We employ Randoop~\cite{pacheco2007randoop}, a widely recognized tool extensively utilized for test case generation. 
Additionally, EvoSuite~\cite{fraser2011evosuite} is executed as a baseline, albeit its primary design for regression testing somewhat constrains its efficacy in triggering defects within the program.
Both Randoop and EvoSuite are allocated a runtime of 3 minutes per defective class, adhering to the methodology outlined in~\cite{shamshiri2015automatically, dinella2022toga}.
We then use scripts from previous works~\cite{dinella2022toga, liu2023towards} to run each test case and check if they trigger any errors.

\textbf{Learning-based Approach.} 
To present the learning-based approach, we employ the whole-test generation model, AthenaTest~\cite{tufano2020unit}.
We also evaluate against TOGA~\cite{dinella2022toga}, a unified transformer-based neural approach to infer both exceptional and assertion test oracles based on the context of the focal method. 
In addition, we fine-tune a seq2seq model, CodeT5+~\cite{wang2023codet5+}, on the dataset used by TOGA to generate unit tests.

\subsection{Evaluation Measures}
\subsubsection{Defect Detection}
To evaluate the effectiveness of \toolname on defect detection, we consider the following metrics: Accuracy, Precision, Recall, F1-score, FPR, and PR-AUC.

\textbf{Accuracy} evaluates the performance that how many methods can be correctly labeled. It is calculated as:
$\frac{TP+ TN}{TP+FP+TN+FN}$.

\textbf{Precision} is the fraction of true defects among the detected ones. It is defined as:
$\frac{TP}{TP+FP}$.

\textbf{Recall} measures how many defects can be correctly detected. It is defined as: 
$\frac{TP}{TP+FN}$.

\textbf{F1-score} is a harmonic mean of $Precision$ and $Recall$ and can be calculated as:
$\frac{2 \times P \times R}{P + R}$.

\textbf{FPR} 
refers to the proportion of non-defects that are predicted to be defects.
It is defined as $\frac{FP}{FP+TN}$.

\textbf{PR-AUC} is the area under the precision-recall curve and is a useful metric of successful prediction when the class distribution is very imbalanced~\cite{hin2022linevd}.

\subsubsection{Unit Test Generation}

To evaluate the effectiveness of \toolname in triggering errors present in the program, the generated unit tests are run on the defective version and the fixed version. 
We consider an error as triggered if a generated unit test fails on the defective version and passes on the fixed version.
Since each fixed version is distinguished from the defective version by a minimal patch fixing the specific error, a test must fail due to the specific error if it only fails on the defective version.
In addition to evaluating the number of errors triggered, we use four metrics defined in~\cite{dinella2022toga}, and we summarize the meaning of these metrics in Table~\ref{tab:metric}.

Following prior work~\cite{liu2023towards}, we also use the Precision metric.
When using error triggering tools, developers care more about the Precision, i.e., how many FPs they need to inspect to trigger an error, and usually have few interests in using a tool with a low Precision~\cite{bessey2010few, johnson2013don}.

\begin{table}[htbp]
\centering
\vspace{-0.1cm}
\caption{The metrics of error triggering}
\vspace{-0.3cm}
\begin{center}
\resizebox{.5\linewidth}{!}{
\begin{tabular}{lcc}
\toprule
\textbf{Metrics} & \textbf{Defective} & \textbf{Fixed} \\
\midrule
True Positive & \textcolor{red}{Fail} & \textcolor{green}{Pass} \\
False Positive & \textcolor{red}{Fail} & \textcolor{red}{Fail} \\
True Negative & \textcolor{green}{Pass} & \textcolor{green}{Pass} \\
False Negative & \textcolor{green}{Pass} & \textcolor{red}{Fail} \\
\bottomrule
\end{tabular}
} 
\vspace{-0.4cm}
\label{tab:metric}
\end{center}
\end{table}

\subsection{Experimental Setting}
We implement \toolname in Python with the help of PyTorch~\cite{pytorch} framework.
All experiments are conducted on an NVIDIA A800 80GB graphics card.
As for defect detection, we utilize \textit{unixcoder-base-nine}~\cite{guo2022unixcoder} from Hugging Face~\cite{huggingface} as our basic model.
We fine-tune \toolname on the studied datasets to obtain a set of suitable parameters.
During the training phase, we use $\mathit{Adam}$ with a batch size of 16 to optimize the parameters of \toolname.
We also leverage $\mathit{GELU}$ as the activation function. 
A dropout of 0.1 is used for dense layers before calculating the final probability. 
We set the maximum number of epochs in our experiment as 200 and adopt an early stop mechanism.
The models (i.e., \toolname and baselines) with the best performance on the validation set are used for the evaluations.
As for unit test generation, we develop the generation pipeline in Python, utilizing PyTorch~\cite{pytorch} implementation of DeepSeek Coder 6.7B and CodeLlama 7B. 
We use the Hugging Face~\cite{huggingface} to load the model weights and generate outputs.
During the attention profiling, we set $\alpha$ to 0.01 and $k$ to 10. 
For each defective method, we generate 100 candidate unit tests (i.e., the candidate number is 100, refer to Section~\ref{sec:rq3} for more details) and test them in the test suite provided by Defects4J.

%% file: sections/results.tex
\section{Results}
To investigate the feasibility of \toolname on defect detection and error triggering, our experiments focus on the following three research questions:

\begin{itemize}[leftmargin=*]
\item \textbf{RQ-1 Comparable Study of Defect Detection.} {\em How well does \toolname perform on method-level defect detection?}

\item \textbf{RQ-2 Comparable Study of Error Triggering.} {\em How well does \toolname perform on triggering the error?}

\item \textbf{RQ-3 Sensitivity Analysis.} {\em How do different configurations affect the overall performance of \toolname?}
\end{itemize}

\input{sections/RQ1}

\input{sections/RQ2}

\input{sections/RQ3}

%% file: sections/RQ1.tex
\subsection{RQ-1 Effectiveness on Defect Detection}
\label{sec:rq1}

\noindent
\textbf{Objective.}
Benefiting from the powerful representation capability of deep neural networks, many DL-based detection approaches have been proposed~\cite{fu2022linevul,ni2023distinguishing}.
CodeBERT and UniXcoder are bimodal pre-trained models for programming languages and natural languages, demonstrating excellent performance across various software engineering tasks, such as code search and code generation~\cite{feng2020codebert, guo2022unixcoder}.
The experiments are conducted to investigate whether \toolname outperforms SOTA method-level detection approaches.

\noindent
\textbf{Experimental Design.}
We consider four SOTA baselines:
LineVul~\cite{fu2022linevul}, SVulD~\cite{ni2023distinguishing}, CodeBERT~\cite{feng2020codebert}, and UniXcoder~\cite{guo2022unixcoder}.
We conduct two distinct experiments to evaluate the performance of \toolname. 
In the first experiment, we undertake the tasks of training, validating, and testing using the Bears and Bugs.jar datasets. 
The second experiment extends our evaluation by training and validating on Bears and Bugs.jar, but testing on the Defects4J dataset.
This experiment aims to showcase the detection capabilities of \toolname in identifying unknown real-world defects.
It is noteworthy that, according to our statistical analysis, there is no overlap in the data extracted from Bears and Bugs.jar with that obtained from Defects4J. 
Our methodology for constructing the training and validation data from Bears and Bugs.jar aligns with established practices in prior research~\cite{fu2022linevul, ni2022defect, ni2023distinguishing}.
Specifically, 80\% of methods are treated as training data, 10\% of methods are treated as validation data, and the left 10\% of methods are treated as testing data.

\noindent
\textbf{Results.}
The evaluation results on Bears and Bugs.jar datasets are reported in Table~\ref{tab:rq1-1} and the best performances are highlighted in bold. 
According to the results, we find that \toolname outperforms all SOTA baseline methods on almost all performance measures except \textit{Recall}. 
In particular, \toolname obtains 0.353, 0.272, and 0.252 in terms of F1-score, Precision, and PR-AUC, which improves baselines by 4.7\% to 7.6\%, 17.7\% to 20.4\%, and 0.4\% to 20.6\% in terms of F1-score, Precision, and PR-AUC, respectively.

\begin{table}[htbp]
  \centering
  \caption{Defect detection results of \toolname compared against four baselines on Bears and Bugs.jar}
  \vspace{-0.2cm}
  \resizebox{.8\linewidth}{!}
  {
    \begin{tabular}{lrrrr}
    \toprule
    \textbf{Methods} & \textbf{F1-score} & \textbf{Recall} & \textbf{Precision} & \textbf{PR-AUC} \\
    \midrule
    LineVul & 0.331 & {0.585} & 0.231  & 0.231 \\
    {SVulD} & 0.337  & \cellcolor{lightgray}\textbf{0.646} & 0.228  & 0.231  \\
    {CodeBERT} & 0.328  & 0.598  & 0.226  & 0.209 \\
    UniXcoder & 0.334 & 0.617 & 0.229 & 0.251 \\
    \midrule
    \cellcolor{lightgray}\textbf{\toolnameb} & \cellcolor{lightgray}\textbf{0.353} & 0.502  &\cellcolor{lightgray} \textbf{0.272} & \cellcolor{lightgray} \textbf{0.252} \\
    \midrule
    \textit{{Improve}} & {\makecell[r]{4.7\% to 7.6\%}} & -- & {\makecell[r]{ 17.7\% to 20.4\%}} & {\makecell[r]{0.4\% to 20.6\%}} \\
    \bottomrule
    \end{tabular}%
    }
    \vspace{-0.1cm}
  \label{tab:rq1-1}%
\end{table}%

To evaluate the defect detection performance of \toolname on unknown real-world Java projects, we train and validate \toolname and baselines on Bears and Bugs.jar datasets. 
Subsequently, we test them on the Defects4J dataset.
The performance comparisons of \toolname and four SOTAs on the Defects4J dataset are presented in Table~\ref{tab:rq1-2}.
According to Table~\ref{tab:rq1-2}, we find that all SOTAs have poor performance on Defects4J dataset, while \toolname outperforms all baselines on almost all performance measures. 
Specifically, \toolname obtains 0.276 of F1-score, 0.198 of Precision, and 0.208 of PR-AUC, which improves baselines by 11.3\% to 35.3\%, 20.0\% to 40.4\%, and 24.6\% to 69.1\%, respectively.
The results indicate that \toolname has a better learning ability than the four baselines.

\begin{table}[htbp]
  \centering
  \caption{Defect detection results of \toolname compared against four baselines on Defects4J\vspace{-0.2cm}}
  
  \resizebox{.8\linewidth}{!}
  {
    \begin{tabular}{lrrrr}
    \toprule
    \textbf{Methods} & \textbf{F1-score} & \textbf{Recall} & \textbf{Precision} & \textbf{PR-AUC} \\
    \midrule
    LineVul & 0.230 & {0.406} & 0.160  & 0.165 \\
    {SVulD} & 0.248  & \cellcolor{lightgray}\textbf{0.504} & 0.164  & 0.167  \\
    {CodeBERT} & 0.204  & 0.369  & 0.141  & 0.123 \\
    UniXcoder & 0.242 & 0.457 & 0.165 & 0.167 \\
    \midrule
    \cellcolor{lightgray}\textbf{\toolnameb} & \cellcolor{lightgray}\textbf{0.276} & 0.453  &\cellcolor{lightgray} \textbf{0.198} & \cellcolor{lightgray} \textbf{0.208} \\
    \midrule
    \textit{{Improve}} & {\makecell[r]{11.3\% to 35.3\%}} & - & {\makecell[r]{20.0\% to 40.4\%}} & {\makecell[r]{24.6\% to 69.1\%}} \\
    \bottomrule
    \end{tabular}%
    }
  \label{tab:rq1-2}%
\end{table}%

In the field of defect detection, the FPR and Accuracy are crucial metrics for assessing detection performance. 
FPR measures the extent of false alarms in the system, indicating the proportion of non-defective samples incorrectly flagged as defective.
On the other hand, Accuracy provides a comprehensive evaluation of the overall correctness of the system.
Therefore, we also compare the performance between \toolname and four baselines in terms of FPR and Accuracy.
According to the results in Table~\ref{tab:rq1-3}, we can observe that \toolname exhibits a reduction in FPR ranging from 14.5\% to 28.7\% compared to other baselines. 
Additionally, its Accuracy shows an improvement ranging from 2.7\% to 5.3\% compared to other baselines. 
Notably, \toolname achieves the highest Accuracy while maintaining the lowest FPR, highlighting its credibility and usability in detecting defects in real-world Java projects.

\begin{table}[!htbp]
  \centering
  \caption{The performance between \toolname and four baselines in terms of FPR and Accuracy}\vspace{-0.2cm}
  \resizebox{.8\linewidth}{!}{
    \begin{tabular}{lrr|rr}
    \toprule
    \textbf{Methods} & \textbf{FPR} & \textbf{\% Decrease} & \textbf{Accuracy} & \textbf{\% Improve} \\
    \midrule
    LineVul & 0.145 & 14.5\% & 0.827 & 2.7\% \\
    {SVulD} & 0.174 &  28.7\% & 0.806 & 5.3\% \\
    {CodeBERT} & 0.152 & 18.4\% & 0.818 & 3.8\% \\
    UniXcoder & 0.157 & 21.0\% & 0.818 & 3.8\% \\
    \midrule
    \rowcolor{lightgray} \textbf{\toolnameb} &  \textbf{0.124} & \textbf{14.5\% to 28.7\%} & \textbf{0.849} & \textbf{2.7\% to 5.3\%} \\
    \bottomrule
    \end{tabular}%
    }
  \vspace{-0.3cm}
  \label{tab:rq1-3}%
\end{table}%

\intuition{
{\bf Answer to RQ-1:}
\toolname outperforms the SOTA baselines at method-level software defect detection.
Specifically, it achieves notable improvements in F1-score, Precision, PR-AUC, and Accuracy, as well as a reduction in FPR.
}

%% file: sections/RQ2.tex
\subsection{RQ-2 Effectiveness on Error Triggering}
\label{sec:rq2}

\noindent
\textbf{Objective.}
In order to enhance LLM's proficiency in generating unit tests that trigger errors, we devised an attention profiling approach.
This involves modifying the attention weights of the LLM to guide its focus toward statements within defective methods that may harbor defects.
This strategic adjustment aims to facilitate the generation of more unit tests that effectively trigger errors.
In this section, our objective is to investigate whether \toolname outperforms previous unit test generation approaches in terms of error triggering.

\noindent
\textbf{Experimental Design.}
To facilitate comparison, we employ DeepSeek Coder and CodeLlama, denoting them as \underline{\toolname and AUGER*}, respectively.
We consider five SOTA baselines:
TOGA~\cite{dinella2022toga}, {EvoSuite}~\cite{fraser2011evosuite}, {Randoop}~\cite{pacheco2007randoop}, AthenaTest~\cite{tufano2020unit}, and CodeT5+~\cite{wang2023codet5+}.

For TOGA, we adhere to the approach outlined in~\cite{liu2023towards}, employing EvoSuite to generate test prefixes on the defective version. 
Subsequently, TOGA generates the corresponding test oracles.
For EvoSuite and Randoop, we employ the scripts provided by the Defects4J toolkit. 
EvoSuite is utilized in its regression mode, while Randoop is applied in both regression and error-revealing modes (i.e., Randoop$_{reg}$ and Randoop$_{rev}$).
For each focal method, we set up \toolname and baselines to repeat 100 times, generating 100 candidate unit tests.
As \toolname requires defect location, we fine-tuned a UniXcoder model to locate the defective statements within the methods using all the defective methods from Bears and Bugs.jar.

We have considered three experimental scenarios.
The first scenario involves conducting experiments using defective methods detected in RQ-1.
The second scenario entails using all defective methods from Defects4J for experimentation.
The third scenario aims to evaluate whether \toolname can trigger errors in real-world projects. 
We follow Defects4J and collect defect-fixing commits from high-quality open-source projects included in Defects4J.
We use DeepSeek Coder for exploration, which collected pre-training data from GitHub before February 2023~\cite{deepseek-coder}.
To prevent data leakage, we only collect defect-fixing commits from March 2023 onwards and obtain 61 defects.
Then we extract methods following the steps in Section~\ref{sec:dataset} and ultimately obtain 41 method-level defects, which are used as input for \toolname to verify its ability to trigger errors in real-world projects.

\noindent
\textbf{Results.}
The effectiveness of \toolname compared against five baselines are reported in Table~\ref{tab:rq2-1}.
According to the results, we can obtain the following observations:

(1) For the defects detected in RQ-1, TOGA, EvoSuite, Randoop$_{reg}$, Randoop$_{rev}$, AUGER*, and \toolname are able to trigger 26, 13, 18, 17, 32, and 35 errors, while AthenaTest and CodeT5+ are unable to trigger any errors.

(2) Similar to (1), for the 723 method-level defects in Defects4J, TOGA, EvoSuite, Randoop$_{reg}$, Randoop$_{rev}$, AUGER*, and \toolname could trigger 61, 26, 39, 35, 78, and 84 errors, respectively, while AthenaTest and CodeT5+ are unable to trigger any errors.
AthenaTest and CodeT5+ generate a large number of candidate unit tests that fail to compile, which makes them extremely ineffective at triggering errors.

(3) Both \toolname and AUGER* show excellent performance, and \toolname works better than AUGER*.
In the next section, we will select the best \toolname to study.
In comparison to the baselines, \toolname triggers 9-35 errors and 23-84 errors more, highlighting the effectiveness of defective information in guiding the LLM. 
In addition, our \toolname also has the highest Precision, meaning that developers only need to inspect a minimum number of unit tests under the same conditions.

\begin{table}[htbp]
  \centering
  \caption{The effectiveness of \toolname compared against baselines on detected defects and all defects\vspace{-0.2cm}}
  \resizebox{\linewidth}{!}{
    \begin{tabular}{l|ccc|ccc}
    \toprule
    \multirow{2}[2]{*}{\textbf{Methods}} & \multicolumn{3}{c|}{\textbf{Detected Defects (420)}} & \multicolumn{3}{c}{\textbf{All Defects (723)}} \\
    \cmidrule{2-7} & \textbf{Trigger} & \textbf{\# Improve} & \textbf{Precision} & \textbf{Trigger} & \textbf{\# Improve} & \textbf{Precision} \\
    \midrule
    TOGA & 26 & 9 & 0.4\% & 61 & 23 & 0.7\% \\
    EvoSuite & 13 & 22 & 0.1\% & 26 & 58 & 0.2\% \\
    Randoop$_{reg}$ & 18 & 17 & 4.6\% & 39 & 45 & 5.6\% \\
    Randoop$_{rev}$ & 17 & 18 & 1.9\% & 35 & 49 & 3.7\% \\
    AthenaTest & 0 & 35 & - & 0 & 84 & - \\
    CodeT5+ & 0 & 35 & - & 0 & 84 & - \\
    \midrule
    \textbf{AUGER*} & 32 & 6-32 & 5.1\% & 78 & 17-78 & 7.9\% \\
    \rowcolor{lightgray} \textbf{\toolnameb} & \textbf{35} & \textbf{9-35} &  \textbf{5.3\%} & \textbf{84} & \textbf{23-84} & \textbf{8.8\%} \\
    \bottomrule
    \end{tabular}%
  }
  \label{tab:rq2-1}%
\end{table}%

To better demonstrate the effectiveness of \toolname, we present the error-triggering quantities for different projects in two scenarios, as shown in Table~\ref{tab:rq2-2}.
According to the results, we can observe that: 
(1) Compared to other projects, \toolname excels at triggering errors in the Chart, Cli, Compress, and Gson projects, achieving a Recall of over 0.15 exceeding 0.15 in these projects.
(2) \toolname performs poorly in projects such as JacksonXml, JxPath, and Mockito, where it fails to trigger any errors.
Upon investigation, we found that most of these defects are multi-hunk defects involving multiple methods.
Triggering these types of defects can be challenging.

\begin{table}[htbp]
  \centering
  \caption{The Recall of \toolname for the projects on Defects4J}\vspace{-0.2cm}
  \resizebox{\linewidth}{!}{
    \begin{tabular}{lrr|lrr}
    \toprule
    \multicolumn{3}{c|}{\textbf{Detected Defects (420)}} & \multicolumn{3}{c}{\textbf{All Defects (723)}} \\
    \midrule
    \textbf{Projects} & \textbf{Recall} & \textbf{Prop.} & \textbf{Projects} & \textbf{Recall} & \textbf{Prop.} \\
    \midrule
    Chart & 0.273 & 3/11 & Chart & 0.240 & 6/25 \\
    Closure & 0 & 0/79 & Closure & 0.030 & 4/134 \\
    Collections & - & 0/0 & Collections & 1 & 1/1 \\
    Csv & 0 & 0/8 & Csv & 0.200 & 3/15 \\
    JacksonCore & 0.077 & 1/13 & JacksonCore & 0.080 & 2/25 \\
    JacksonXml & 0 & 0/3 & JacksonXml & 0 & 0/5 \\
    JxPath & 0 & 0/13 & JxPath & 0 & 0/21 \\
    Math & 0.139 & 10/72 & Math & 0.220 & 22/100 \\
    Time & 0 & 0/11 & Time & 0.087 & 2/23 \\
    Cli & 0.300 & 6/20 & Cli & 0.162 & 6/37 \\
    Codec & 0.111 & 1/9 & Codec & 0.250 & 4/16 \\
    Compress & 0.231 & 6/26 & Compress & 0.196 & 9/46 \\
    Gson & 0.200 & 2/10 & Gson & 0.188 & 3/16 \\
    JacksonDatabind & 0 & 0/62 & JacksonDatabind & 0.031 & 3/96 \\
    Jsoup & 0.051 & 2/39 & Jsoup & 0.178 & 13/73 \\
    Lang & 0.121 & 4/33 & Lang & 0.107 & 6/56 \\
    Mockito & 0 & 0/11 & Mockito & 0 & 0/34 \\
    \midrule
    \rowcolor{lightgray} \textbf{Sum} & \textbf{0.083} & \textbf{35/420} & \textbf{Sum} & \textbf{0.116} & \textbf{84/723} \\
    \bottomrule
    \end{tabular}%
  }
  \label{tab:rq2-2}%
\end{table}%

\begin{figure}[htbp]
    \centering
    \vspace{-0.2cm}
    \includegraphics[width=.6\linewidth]{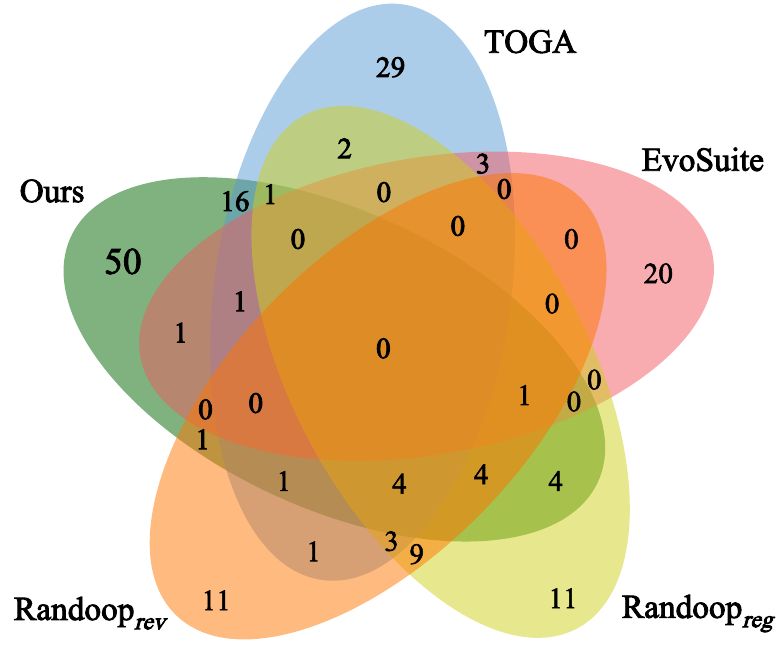}
    \caption{Venn diagram of \toolname and studied baselines}
    \vspace{-0.2cm}
    \label{fig:venn}
\end{figure}

We draw a Venn diagram to further illustrate the performance difference on error triggering.
For a better presentation, we independently illustrate the Top-4 best baselines (i.e., TOGA, EvoSuite, Randoop$_{reg}$, and Randoop$_{rev}$) on the basis of the number of triggering errors and ignore the baselines (i.e., AthenaTest and CodeT5+) that cannot trigger error for easy reference.
Fig.~\ref{fig:venn} shows the illustrated results and we can also obtain two observations: 
(1) Individual approaches have unique abilities to trigger specific errors that others cannot, making their performance somewhat complementary.
(2) Overall, \toolname has a more powerful ability than baselines since it can trigger the most number of unique errors (i.e., 50) that other baselines can hardly trigger.

\begin{figure}[htbp]
    \centering
     \vspace{-0.2cm}
     \includegraphics[width=\linewidth]{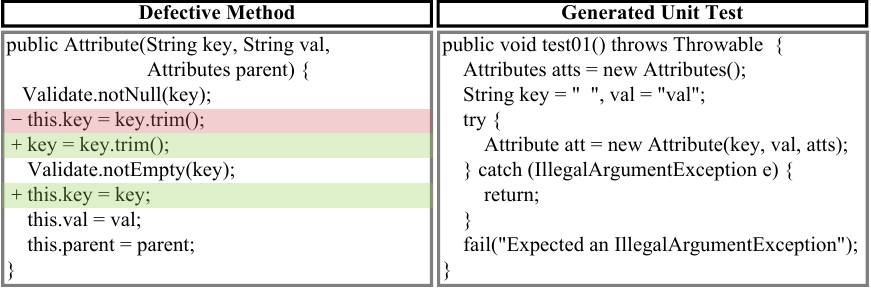}
    \caption{Unique error triggered by \toolname}
     \vspace{-0.2cm}
    \label{fig:unique}
\end{figure}

\underline{{\bf Case Study.}} 
To explore why \toolname has an outstanding performance in triggering unique errors, we further analyze one example (i.e., \href{https://github.com/jhy/jsoup/issues/1159}{Jsoup-85}) as a case study, as shown in Fig.~\ref{fig:unique}. 
The defect is that if key itself contains only space characters, then \textit{key.trim()} will get an empty string.
After the empty string is assigned to \textit{this.key}, the \textit{Validate.notEmpty(key)} check passes because key itself is not an empty string.
But in reality, the value of \textit{this.key} is an empty string, which is an incorrect state.
Existing approaches cannot understand the defect present in the code, making it difficult to efficiently generate unit tests that trigger the error.
As shown in Fig.~\ref{fig:unique}, \toolname understands the semantics of the code correctly and generates a unit test that effectively exposes the defect in the defective method when handling key that only contain spaces.
This example further exemplifies the capability of \toolname to leverage defective information to guide LLM to generate error-triggering unit tests efficiently.

\textbf{\underline{Effectiveness of \toolnameb in real-world projects.}} 
Since the DeepSeek Coder's pre-training data was sourced from GitHub prior to February 2023, in order to prevent data leakage, we collect defects from real-world projects starting from March 2023 to evaluate \toolname's error-triggering capabilities. 
Table~\ref{tab:rq2-3} presents the results of \toolname for these real-world projects. 
According to the results, we can conclude that \toolname possesses the ability to trigger errors in real-world projects, not just limited to the defects present in the Defects4J dataset. 
This further demonstrates the practicality and feasibility of \toolname in generating unit tests.
\toolname has the ability to guide LLM in generating effective unit tests, triggering potential errors in future real-world projects.

\begin{table}[htbp]
\centering
\caption{The Recall of \toolname for the real-world projects}
\resizebox{.8\linewidth}{!}
{
    \begin{tabular}{lrr|lrr}
    \toprule
    \textbf{Project} & \textbf{Recall} & \textbf{Prop.} & \multicolumn{1}{r}{\textbf{Project}} & \textbf{Recall} & \multicolumn{1}{r}{\textbf{Prop.}} \\
    \midrule
    Cli & 0 & 0/3 & Jsoup & 0.056 & 1/18 \\
    Codec & 0.250 & 1/4 & Lang & 0.375 & 3/8 \\
    Compress & 0.125 & 1/8 & \cellcolor{lightgray}\textbf{Sum} & \cellcolor{lightgray}\textbf{0.146} & \cellcolor{lightgray}\textbf{6/41} \\
    \bottomrule  
    \end{tabular}%
}
 \vspace{-0.3cm}
\label{tab:rq2-3}%
\end{table}%

\intuition{
{\bf Answer to RQ-2}: 
\toolname achieves better performance in both the defects detected in RQ-1 and all method-level defects in Defects4J, triggering 35 and 84 errors, respectively.
\toolname can also trigger potential errors in future real-world projects.
}

%% file: sections/RQ3.tex
\subsection{RQ-3 Configurations of \toolname}
\label{sec:rq3}

\noindent
\textbf{Objective.}
In defect detection, we have incorporated \textit{Adversarial Learning} and \textit{Contrastive Learning} to enhance the performance and robustness of the UniXcoder.
As a result, we aim to investigate the individual effects of these two components.
In error triggering, we modify attention to guide LLM focus on defective statements, generating unit tests that trigger errors.
During this process, we seek to examine the impact of attention modifying, as well as the influence of the candidate number of unit tests on the final results.

\noindent
\textbf{Experimental Design.}
First, we investigate the impact of different components on defect detection and design three variants of \toolname. 
\toolname-v1 represents the removal of \textit{Adversarial Learning} and \textit{Contrastive Learning}, \toolname-v2 indicates the removal of \textit{Adversarial Learning}, and \toolname-v3 signifies the removal of \textit{Contrastive Learning}. 
This approach allows us to examine the individual effects of each component.
Subsequently, we explore the influence of attention modifying and defect location on error triggering, creating two variants.
AUGER$_{w/o}$ denotes the elimination of the attention-modifying component, while AUGER$_{gt}$ denotes the utilization of Ground-Truth defect location to guide the LLM.
Finally, we conduct a comparative analysis of the impact of candidate numbers on \toolname and TOGA, the latter of which is identified as the baseline with the best performance (cf. Section~\ref{sec:rq2}).

\noindent
\textbf{Results.} We discuss the results from the aspects of ablation and unit test candidate number, respectively.

\begin{table}[htbp]
  \centering
  \caption{Defect detection results of \toolname compared against variants\vspace{-0.2cm}}
  \resizebox{.8\linewidth}{!}
  {
    \begin{tabular}{lrrrrrr}
    \toprule
    \textbf{Methods} & \textbf{AL} & \textbf{CL} & \textbf{F1-score} & \textbf{PR-AUC} & \textbf{FPR} \\
    \midrule
    \toolname-v1 & \ding{55} & \ding{55} & 0.242 & 0.167 & 0.157 \\
    \toolname-v2 & \ding{55} & \ding{51} & 0.248  & 0.167 & 0.174 \\
    \toolname-v3 & \ding{51} & \ding{55} & 0.261  & 0.180 & 0.143 \\
    \midrule
    \rowcolor{lightgray}
    \textbf{\toolnameb} & \ding{51} & \ding{51} & \textbf{0.276} & \textbf{0.208} & \textbf{0.124} \\
    \bottomrule
    \end{tabular}%
    }
  \label{tab:rq3-1}%
\end{table}%

\underline{\textbf{Impact of components in defect detection.}}
Table~\ref{tab:rq3-1} shows the effectiveness of different variants and the better performance is highlighted in bold. 
According to the results, we can observe that:
(1) Two components have their own advantages in a method-level defect detection scenario, achieving a varying performance and significantly improving the performance of \toolname-v1. 
Both of them can contribute to the performance of \toolname. 
(2) \textit{Adversarial Learning} demonstrates superior performance compared to \textit{Contrastive Learning}. 
Specifically, \toolname-v3 outperforms \toolname-v2 in terms of F1-score (0.248$\rightarrow$0.261), PR-AUC (0.167$\rightarrow$0.180), and FPR (0.174$\rightarrow$0.143) metrics.
(3) A combination of these two components yields optimal performance in terms of F1-score (0.276), PR-AUC (0.208), and FPR (0.124). 
This suggests that incorporating \textit{Adversarial Learning} and \textit{Contrastive Learning} can enhance the effectiveness of the pre-trained model for defect detection.

\begin{table}[htbp]
    \centering
    \caption{Error triggering results of \toolname compared against variants\vspace{-0.2cm}}
    \resizebox{.8\linewidth}{!}
    {
    \begin{threeparttable}   
        \begin{tabular}{lccc}
        \toprule
        \textbf{Datasets} & \textbf{AUGER$_{w/o}$} & \textbf{AUGER} & \textbf{AUGER$_{gt}$} \\
        \midrule
        {Detected Defects} & {28} & 35 & \textbf{44} \\
        {All Defects} & {67} & 84 & \textbf{99} \\
        \bottomrule
        \end{tabular}
        \end{threeparttable}
    }
    \label{tab:selection}
\end{table}

\underline{\textbf{Impact of components in error triggering.}}
According to the results in Table~\ref{tab:selection}, we observe that: 
(1) \toolname and AUGER$_{gt}$ trigger more errors than AUGER$_{w/o}$, which indicates the importance of the attention modifying. 
Particularly, \toolname and AUGER$_{gt}$ can trigger 17 and 32 more errors than AUGER$_{w/o}$ in all defects.
(2) \toolname achieves comparable performance, obtaining similar conclusions in two different scenarios: weaker than AUGER$_{gt}$ but stronger than AUGER$_{w/o}$. 
As expected, AUGER$_{gt}$ performs the best, as it utilizes Ground-Truth defect location information to guide the LLM, avoiding the misguidance caused by errors in defect location.

\begin{figure}[htbp]
    \centering
    \includegraphics[width=.77\linewidth]{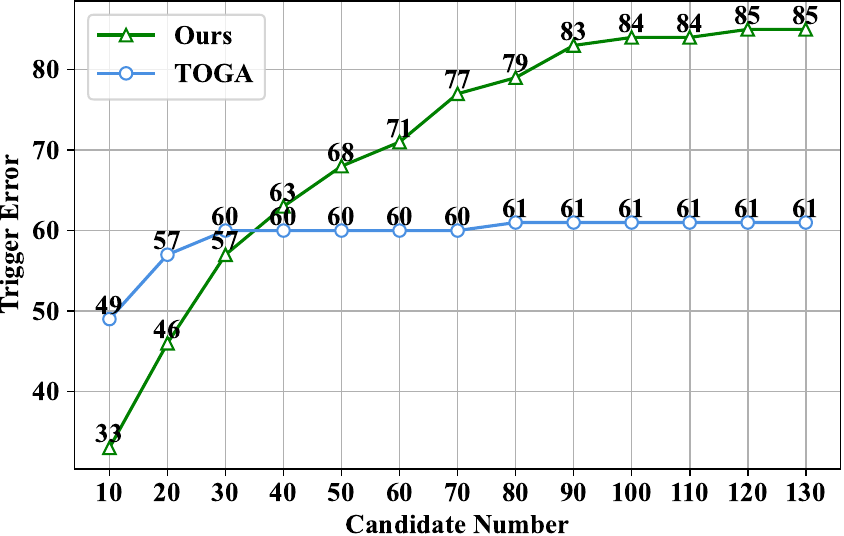}
    \caption{The varying performance of \toolname and TOGA with different unit test candidate number on all defects}
    \label{fig:candidate}
\end{figure}

\underline{\textbf{Impact of unit test candidate number.}}
According to the results in Fig.~\ref{fig:candidate}, we find that: 
(1) Different candidate numbers have varying impacts on \toolname's performance and the performance of \toolname increases as the number of candidates increases.
(2) Through the improvement curve, it can be observed that the performance improvement of \toolname far exceeds that of TOGA. 
As the number of candidate unit tests increases, the improvement curve of TOGA remains relatively flat, reaching optimal effectiveness when generating 80 candidate unit tests. 
In contrast, \toolname exhibits an upward trend, reaching optimal effectiveness when generating 120 candidate unit tests.
(3) More candidate numbers may not guarantee additional performance improvement.
When \toolname generates 70 candidate unit tests, there is a significant improvement compared to generating 10 candidate unit tests (i.e., 33$\rightarrow$77). 
However, when continuously increasing the number of candidates, the rate of performance improvement decreases (i.e., 77$\rightarrow$85) and meanwhile, the generation cost with LLM is increasing.
Considering both the performance improvement and the generation cost caused by LLM, we adopt 100 unit test candidate numbers as the default setting.

\intuition{
{\bf Answer to RQ-3}: 
(1) The two components (i.e., Adversarial Learning and Contrastive Learning) contribute substantially to \toolname, and combining them achieves the best performance of defect detection.
(2) Attention modifying can direct LLM to focus on defective statements to generate more unit tests that trigger errors.
(3) Increasing the candidate number can significantly improve the error-triggering performance of \toolname, but will gradually become saturated.
}

%% file: sections/threats.tex
\section{Discussion}

\subsection{Comparison of Efficiency}

\noindent
\textbf{Defect Detection.}
Table~\ref{tab:efficiency_detection} presents the details of time cost and GPU memory cost for baselines and \toolname. 
We find that although \toolname requires more time and computational resources during fine-tuning, its inference costs are comparable to those of the baselines.
On the whole, during fine-tuning, \toolname takes an average of 3 minutes and 52 seconds per epoch and consumes 24,538M of GPU memory. 
In contrast, the baselines take an average of 1 minute and 42 seconds to 3 minutes and 24 seconds per epoch, with a GPU memory cost ranging from 13,412M to 23,442M. 
However, \toolname only requires 16 seconds and 5,658M of GPU memory during inference, which is very close to the baselines (i.e., 16 seconds for time cost and 4,626M to 5,754M for GPU memory cost).
Given the performance improvements of \toolname (refer to Section~\ref{sec:rq1}), the additional resources spent on fine-tuning are justified. 
Furthermore, \toolname does not require additional resources for inference after fine-tuning.

\begin{table}[htbp]
    \centering
    \caption{The time cost and GPU memory cost of baselines and \toolname for one epoch on defect detection}
    \vspace{-0.2cm}
    \resizebox{.8\linewidth}{!}
    {
    \begin{threeparttable}   
        \begin{tabular}{lcccc}
        \toprule
        \multirow{2.5}{*}{\textbf{Methods}} & \multicolumn{2}{c}{\textbf{Time Cost}} & \multicolumn{2}{c}{\textbf{GPU Memory Cost}} \\
        \cmidrule{2-5}
        & \multicolumn{1}{c}{\textbf{Fine-Tuning}} & \multicolumn{1}{c}{\textbf{Inference}} & \multicolumn{1}{c}{\textbf{Fine-Tuning}} & \multicolumn{1}{c}{\textbf{Inference}} \\
        \midrule
        LineVul & 1m 16s & 16s & 13,924M & 4,626M \\
        SVulD & 3m 24s & 16s & 23,442M & 5,754M \\
        CodeBERT & 1m 42s & 16s & 13,412M & 5,650M \\
        UniXcoder & 1m 45s & 16s & 13,416M & 5,658M \\
        \midrule
        \toolname & 3m 52s & 16s & 24,538M & 5,658M \\
        \bottomrule
        \end{tabular}
        \end{threeparttable}
    }
    \label{tab:efficiency_detection}
\end{table}

\begin{figure}[htbp]
    \vspace{-0.2cm}
    \centering
    \includegraphics[width=.8\linewidth]{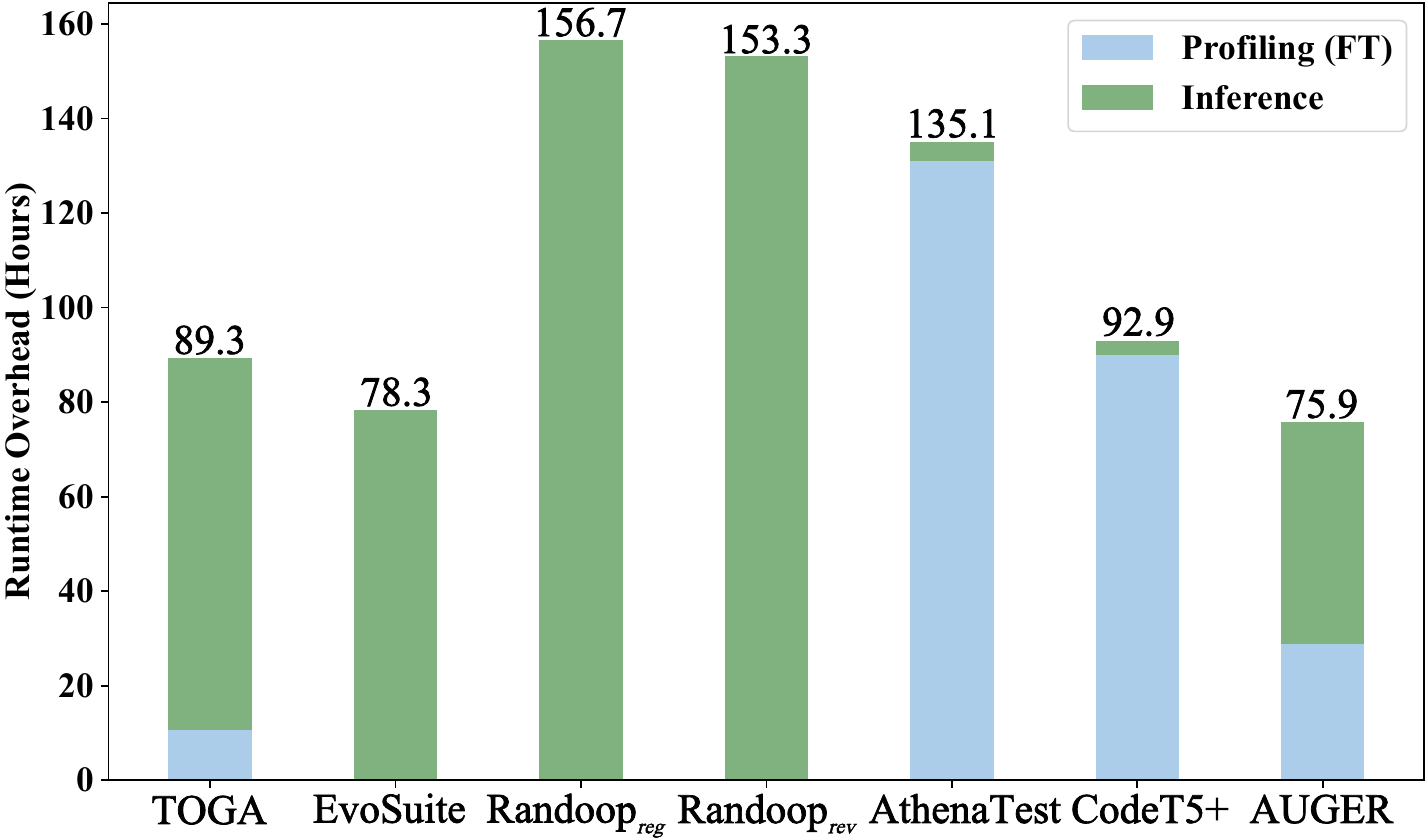}
    \caption{The runtime overheads of baselines and \toolname}
    \label{fig:efficiency_generation}
    \vspace{-0.5cm}
\end{figure}

\noindent
\textbf{Unit Test Generation.}
Fig.~\ref{fig:efficiency_generation} shows the average runtime used by each baseline and \toolname on unit test generation.
For TOGA, AthenaTest, and CodeT5+, we record the total fine-tuning time (i.e., FT in Fig.~\ref{fig:efficiency_generation}) and inference time during our experiments. 
TOGA uses the default settings from previous work~\cite{liu2023towards}, while AthenaTest and CodeT5+ undergo 10 epochs of fine-tuning.
For EvoSuite, Randoop$_{reg}$, and Randoop$_{rev}$, which do not require fine-tuning and instead directly use scripts provided by Defects4J to generate test cases, we only record the test case generation time (i.e., Inference in Fig.~\ref{fig:efficiency_generation}).
Our \toolname does not require training, so we record the time for the attention profiling process (i.e., Profiling in Fig.~\ref{fig:efficiency_generation}) and the test case generation time.
Obviously, we find that the total time required for \toolname is less than that of other baselines.
Even though \toolname involves both profiling and inference processes, the time consumption for these processes is relatively low. 
For example, \toolname requires a total of 75.9 hours, whereas the baselines require between 78.3 hours and 156.7 hours.
Although AthenaTest and CodeT5+ require minimal inference time, they require significantly longer fine-tuning times and ultimately perform poorly in error triggering (refer to Section~\ref{sec:rq2}).
Overall, \toolname uses the least total time and achieves the best error trigger results, making it more effective in real-world scenarios.

\subsection{Threats to Validity}

\noindent
\textbf{Internal Validity.}
The internal threat arises from potential data leakage since referenced unit tests may be part of the training data of LLM.
To tackle this issue, we initially calculate the number of error-triggering unit tests generated by \toolname, which matches the reference unit test in Defects4J.
We find that out of 84 triggered errors, 0 of error-triggering unit tests align with the unit tests in the fixed version. 
Only 12 unit tests are similar but not identical (e.g., input and output are different) to the unit tests in the fixed version.
Additionally, compared to the basic LLM (i.e., DeepSeek Coder), \toolname demonstrates a significant enhancement in performance, triggering 17 more errors.
This demonstrates that the improved results achieved by \toolname are not merely a result of memorizing the training data.
Moreover, the pre-training data for DeepSeek Coder was collected from GitHub before February 2023~\cite{deepseek-coder}. 
To prevent data leakage, we also collected defects from real-world projects starting from March 2023 to evaluate \toolname's error-triggering capabilities.
According to the results, \toolname can trigger 6 out of 41 errors. Therefore, we can conclude that \toolname possesses the ability to reveal potential errors in future real-world projects.
The second internal threat arises from the reliance of unit test generation on defect detection.
In this experiment, we conducted software defect detection on widely used datasets (e.g., Defects4J), and our model achieved the lowest FPR (i.e., 0.124) and the highest F1-score (i.e., 0.276), thereby mitigating the internal threat to the experimental design.

\noindent
\textbf{External Validity.}
The effectiveness observed in \toolname's performance may not be applicable across different datasets. 
We conduct evaluations not only on the widely-used Defects4J dataset but also on the collected dataset from real-world projects. 
This broader evaluation scope aims to showcase the generalizability of our approach.

%% file: sections/related_work.tex
\section{Related Work}

\subsection{Defect Detection}
Traditional work on defect detection has explored a broad spectrum of metrics, encompassing factors such as code size~\cite{menzies2006data}, code complexity~\cite{zimmermann2007predicting}, object-oriented~\cite{menzies2006data}, organizational~\cite{nagappan2008influence}, and change history~\cite{nagappan2005use}), to forecast potential defects in software projects.
Graves et al.~\cite{graves2000predicting} introduced the idea that recent modifications to code serve as effective indicators of impending defects, while Kim et al.~\cite{kim2007predicting} noted that defects often manifest in clusters within the history of software changes, meaning that recent changes and faults are likely to introduce defects in the future.
DL-based approaches have been proposed to learn from historical data~\cite{zhou2019devign, hin2022linevd, li2021vulnerability, chakraborty2021deep, li2019deepfl, fu2022linevul, wardat2021deeplocalize, yamaguchi2014modeling, li2018vuldeepecker, li2021vuldeelocator}.
CodeBERT~\cite{feng2020codebert} and UniXcoder~\cite{guo2022unixcoder} are bimodal pre-trained models for programming languages and natural languages.
They learn general representations that support downstream applications such as vulnerability detection~\cite{hin2022linevd}, defect prediction~\cite{ni2022best}, etc.
Ni et al.~\cite{ni2023distinguishing} proposed SVulD by adopting contrastive learning to train the UniXcoder model for learning distinguishing semantic representation of functions regardless of their lexically similar information.
However, these approaches often provide solely binary predictions, indicating whether a code snippet or statement contains defects or not~\cite{li2019deepfl, yang2024large, gao2018cobot}.
To address this limitation, Fu et al. proposed LineVul~\cite{fu2022linevul}, which uses attention scores to elucidate the reasoning behind the model's decisions. 
Additionally, several approaches improve the model's performance by integrating the graph structure of the code~\cite{zhou2019devign, hin2022linevd, li2021vulnerability, chakraborty2021deep, lou2021boosting, li2022fault}.
Just-In-Time (JIT) defect prediction approaches~\cite{zhao2023systematic, misirli2016studying,pornprasit2021jitline} have been proposed to predict whether a commit will introduce defects in the future. 
PyExplainer~\cite{pornprasit2021pyexplainer} is a novel, local, rule-based, model-agnostic technique designed to explain the predictions of JIT defect models.
While these efforts have incorporated explanatory features, they often focus on why the model made a specific decision rather than providing detailed insights into the conditions that cause defects (e.g., the inputs and outputs that can expose defects).

Different from previous works, our paper connects defect detection and unit test generation. 
For each detected defect, \toolname generates corresponding test cases that trigger the error, instilling greater confidence in the defect detection results.

\vspace{-0.3cm}
\subsection{Unit Test Generation}
Unit test generation approaches can be classified into two types: traditional generation approaches and learning-based generation approaches. 
However, both types face challenges related to inefficiency.
Traditional generation approaches~\cite{fraser2011evosuite, pacheco2007randoop} focus on code coverage, and research shows that traditional approaches are very effective at achieving high coverage~\cite{aleti2017analysing, oliveira2018mapping, panichella2015reformulating, panichella2017automated}.
Randoop~\cite{pacheco2007randoop} is a widely recognized tool extensively utilized for generating unit tests for Java code using feedback-directed random test generation.
EvoSuite is a tool that automates the generation of test suites, aiming for high code coverage, minimal size, and comprehensive assertions. 
However, previous studies show that high code coverage does not necessarily imply effective error triggering~\cite{almasi2017industrial, shamshiri2015automatically}.
Recently, learning-based generation approaches~\cite{alagarsamy2023a3test, tufano2020unit, dinella2022toga} have achieved significant progress.
AthenaTest~\cite{tufano2020unit} is a Transformer-based model that is learned from developer-written test cases in order to generate correct and readable tests. 
The task is framed as a translation problem, where the source is a focal method and the target is the corresponding test case.
TOGA~\cite{dinella2022toga} is a unified transformer-based neural approach to infer both exceptional and assertion test oracles based on the context of the focal method.
A3Test~\cite{alagarsamy2023a3test} leverages the domain adaptation principles where the goal is to adapt the existing knowledge from an assertion generation task to the test case generation task.
However, these approaches focus on randomly generating a large number of test cases by fitting the Method2Test dataset~\cite{tufano2022methods2test}, which contains numerous clean methods and non-error-triggering test cases, lacking effective information guidance.
This results in inefficient tests and an inability to efficiently trigger errors.

Different from existing works, our paper focuses on employing defect location information to guide the generation of unit tests, reducing the model’s search space and enhancing the efficiency of unit test generation.

%% file: sections/conclusion.tex
\section{Conclusion and Future Work}
We propose \toolname, which is a method-level approach for defect detection and error triggering.
\toolname first detects the proneness of defects and then guides the LLM to generate unit tests for triggering such an error with the help of critical information present in defective code.
As a result, \toolname will provide developers with defect detection results, convincing error-triggering unit tests, and corresponding test results. 
This automated approach instills confidence in developers regarding defect detection results.
To evaluate the effectiveness of \toolname, we conduct a large-scale experiment by comparing it with SOTAs on the widely used dataset Defects4J.
\toolname makes great improvements by 11.3\% to 35.3\%, 20.0\% to 40.4\%, and 24.6\% to 69.1\% in terms of F1-score, Precision, and PR-AUC, and can trigger 23 to 84 more bugs than SOTAs.
Besides, we also conduct a further study to verify the generalization in practical usage by collecting a new dataset from real-world projects.
In the future, we will generate multi-dimensional reports on the test results and conduct human studies to fully validate the effectiveness of \toolname.

%% file: main.bbl
\begin{thebibliography}{10}
\providecommand{\url}[1]{#1}
\csname url@samestyle\endcsname
\providecommand{\newblock}{\relax}
\providecommand{\bibinfo}[2]{#2}
\providecommand{\BIBentrySTDinterwordspacing}{\spaceskip=0pt\relax}
\providecommand{\BIBentryALTinterwordstretchfactor}{4}
\providecommand{\BIBentryALTinterwordspacing}{\spaceskip=\fontdimen2\font plus
\BIBentryALTinterwordstretchfactor\fontdimen3\font minus \fontdimen4\font\relax}
\providecommand{\BIBforeignlanguage}[2]{{%
\expandafter\ifx\csname l@#1\endcsname\relax
\typeout{** WARNING: IEEEtran.bst: No hyphenation pattern has been}%
\typeout{** loaded for the language `#1'. Using the pattern for}%
\typeout{** the default language instead.}%
\else
\language=\csname l@#1\endcsname
\fi
#2}}
\providecommand{\BIBdecl}{\relax}
\BIBdecl

\bibitem{wardat2021deeplocalize}
M.~Wardat, W.~Le, and H.~Rajan, ``Deeplocalize: Fault localization for deep neural networks,'' in \emph{2021 IEEE/ACM 43rd International Conference on Software Engineering (ICSE)}.\hskip 1em plus 0.5em minus 0.4em\relax IEEE, 2021, pp. 251--262.

\bibitem{yamaguchi2014modeling}
F.~Yamaguchi, N.~Golde, D.~Arp, and K.~Rieck, ``Modeling and discovering vulnerabilities with code property graphs,'' in \emph{2014 IEEE Symposium on Security and Privacy}.\hskip 1em plus 0.5em minus 0.4em\relax IEEE, 2014, pp. 590--604.

\bibitem{li2018vuldeepecker}
Z.~Li, D.~Zou, S.~Xu, X.~Ou, H.~Jin, S.~Wang, Z.~Deng, and Y.~Zhong, ``Vuldeepecker: A deep learning-based system for vulnerability detection,'' in \emph{Proceedings of the 25th Annual Network and Distributed System Security Symposium}, 2018.

\bibitem{li2021vuldeelocator}
Z.~Li, D.~Zou, S.~Xu, Z.~Chen, Y.~Zhu, and H.~Jin, ``Vuldeelocator: a deep learning-based fine-grained vulnerability detector,'' \emph{IEEE Transactions on Dependable and Secure Computing}, 2021.

\bibitem{yang2024large}
A.~Z. Yang, C.~Le~Goues, R.~Martins, and V.~Hellendoorn, ``Large language models for test-free fault localization,'' in \emph{Proceedings of the 46th IEEE/ACM International Conference on Software Engineering}, 2024, pp. 1--12.

\bibitem{gao2018cobot}
Q.~Gao, S.~Ma, S.~Shao, Y.~Sui, G.~Zhao, L.~Ma, X.~Ma, F.~Duan, X.~Deng, S.~Zhang \emph{et~al.}, ``Cobot: static c/c++ bug detection in the presence of incomplete code,'' in \emph{Proceedings of the 26th Conference on Program Comprehension}, 2018, pp. 385--388.

\bibitem{li2019deepfl}
X.~Li, W.~Li, Y.~Zhang, and L.~Zhang, ``Deepfl: Integrating multiple fault diagnosis dimensions for deep fault localization,'' in \emph{Proceedings of the 28th ACM SIGSOFT international symposium on software testing and analysis}, 2019, pp. 169--180.

\bibitem{yin2024multitask}
X.~Yin, C.~Ni, and S.~Wang, ``Multitask-based evaluation of open-source llm on software vulnerability,'' \emph{IEEE Transactions on Software Engineering}, 2024.

\bibitem{fu2022linevul}
M.~Fu and C.~Tantithamthavorn, ``Linevul: A transformer-based line-level vulnerability prediction,'' in \emph{Proceedings of the 19th International Conference on Mining Software Repositories}, 2022, pp. 608--620.

\bibitem{zhou2019devign}
Y.~Zhou, S.~Liu, J.~Siow, X.~Du, and Y.~Liu, ``Devign: Effective vulnerability identification by learning comprehensive program semantics via graph neural networks,'' in \emph{In Proceedings of the 33rd International Conference on Neural Information Processing Systems}, 2019, p. 10197–10207.

\bibitem{hin2022linevd}
D.~Hin, A.~Kan, H.~Chen, and M.~A. Babar, ``Linevd: statement-level vulnerability detection using graph neural networks,'' in \emph{Proceedings of the 19th international conference on mining software repositories}, 2022, pp. 596--607.

\bibitem{li2021vulnerability}
Y.~Li, S.~Wang, and T.~N. Nguyen, ``Vulnerability detection with fine-grained interpretations,'' in \emph{Proceedings of the 29th ACM Joint Meeting on European Software Engineering Conference and Symposium on the Foundations of Software Engineering}, 2021, pp. 292--303.

\bibitem{chakraborty2021deep}
S.~Chakraborty, R.~Krishna, Y.~Ding, and B.~Ray, ``Deep learning based vulnerability detection: Are we there yet,'' \emph{IEEE Transactions on Software Engineering}, 2021.

\bibitem{lou2021boosting}
Y.~Lou, Q.~Zhu, J.~Dong, X.~Li, Z.~Sun, D.~Hao, L.~Zhang, and L.~Zhang, ``Boosting coverage-based fault localization via graph-based representation learning,'' in \emph{Proceedings of the 29th ACM Joint Meeting on European Software Engineering Conference and Symposium on the Foundations of Software Engineering}, 2021, pp. 664--676.

\bibitem{li2022fault}
Y.~Li, S.~Wang, and T.~N. Nguyen, ``Fault localization to detect co-change fixing locations,'' in \emph{Proceedings of the 30th ACM Joint European Software Engineering Conference and Symposium on the Foundations of Software Engineering}, 2022, pp. 659--671.

\bibitem{steenhoek2023empirical}
B.~Steenhoek, M.~M. Rahman, R.~Jiles, and W.~Le, ``An empirical study of deep learning models for vulnerability detection,'' in \emph{2023 IEEE/ACM 45th International Conference on Software Engineering (ICSE)}.\hskip 1em plus 0.5em minus 0.4em\relax IEEE, 2023, pp. 2237--2248.

\bibitem{almasi2017industrial}
M.~M. Almasi, H.~Hemmati, G.~Fraser, A.~Arcuri, and J.~Benefelds, ``An industrial evaluation of unit test generation: Finding real faults in a financial application,'' in \emph{2017 IEEE/ACM 39th International Conference on Software Engineering: Software Engineering in Practice Track (ICSE-SEIP)}.\hskip 1em plus 0.5em minus 0.4em\relax IEEE, 2017, pp. 263--272.

\bibitem{shamshiri2015automatically}
S.~Shamshiri, R.~Just, J.~M. Rojas, G.~Fraser, P.~McMinn, and A.~Arcuri, ``Do automatically generated unit tests find real faults? an empirical study of effectiveness and challenges (t),'' in \emph{2015 30th IEEE/ACM International Conference on Automated Software Engineering (ASE)}.\hskip 1em plus 0.5em minus 0.4em\relax IEEE, 2015, pp. 201--211.

\bibitem{alagarsamy2023a3test}
S.~Alagarsamy, C.~Tantithamthavorn, and A.~Aleti, ``A3test: Assertion-augmented automated test case generation,'' \emph{arXiv preprint arXiv:2302.10352}, 2023.

\bibitem{tufano2020unit}
M.~Tufano, D.~Drain, A.~Svyatkovskiy, S.~K. Deng, and N.~Sundaresan, ``Unit test case generation with transformers and focal context,'' \emph{arXiv preprint arXiv:2009.05617}, 2020.

\bibitem{dinella2022toga}
E.~Dinella, G.~Ryan, T.~Mytkowicz, and S.~K. Lahiri, ``Toga: A neural method for test oracle generation,'' in \emph{Proceedings of the 44th International Conference on Software Engineering}, 2022, pp. 2130--2141.

\bibitem{tufano2022methods2test}
M.~Tufano, S.~K. Deng, N.~Sundaresan, and A.~Svyatkovskiy, ``Methods2test: A dataset of focal methods mapped to test cases,'' in \emph{Proceedings of the 19th International Conference on Mining Software Repositories}, 2022, pp. 299--303.

\bibitem{madeiral2019bears}
F.~Madeiral, S.~Urli, M.~Maia, and M.~Monperrus, ``Bears: An extensible java bug benchmark for automatic program repair studies,'' in \emph{2019 IEEE 26th International Conference on Software Analysis, Evolution and Reengineering (SANER)}.\hskip 1em plus 0.5em minus 0.4em\relax IEEE, 2019, pp. 468--478.

\bibitem{saha2018bugs}
R.~K. Saha, Y.~Lyu, W.~Lam, H.~Yoshida, and M.~R. Prasad, ``Bugs. jar: A large-scale, diverse dataset of real-world java bugs,'' in \emph{Proceedings of the 15th international conference on mining software repositories}, 2018, pp. 10--13.

\bibitem{just2014defects4j}
R.~Just, D.~Jalali, and M.~D. Ernst, ``Defects4j: A database of existing faults to enable controlled testing studies for java programs,'' in \emph{Proceedings of the 2014 international symposium on software testing and analysis}, 2014, pp. 437--440.

\bibitem{replication}
\BIBentryALTinterwordspacing
``Replication,'' 2024. [Online]. Available: \url{https://github.com/vinci-grape/AUGER}
\BIBentrySTDinterwordspacing

\bibitem{ni2024learning}
C.~Ni, L.~Shen, X.~Xu, X.~Yin, and S.~Wang, ``Learning-based models for vulnerability detection: An extensive study,'' \emph{arXiv preprint arXiv:2408.07526}, 2024.

\bibitem{ni2023distinguishing}
C.~Ni, X.~Yin, K.~Yang, D.~Zhao, Z.~Xing, and X.~Xia, ``Distinguishing look-alike innocent and vulnerable code by subtle semantic representation learning and explanation,'' in \emph{Proceedings of the 31st ACM Joint European Software Engineering Conference and Symposium on the Foundations of Software Engineering}, 2023, pp. 1611--1622.

\bibitem{guo2022unixcoder}
D.~Guo, S.~Lu, N.~Duan, Y.~Wang, M.~Zhou, and J.~Yin, ``Unixcoder: Unified cross-modal pre-training for code representation,'' in \emph{Proceedings of the 60th Annual Meeting of the Association for Computational Linguistics (Volume 1: Long Papers)}, 2022, pp. 7212--7225.

\bibitem{feng2020codebert}
Z.~Feng, D.~Guo, D.~Tang, N.~Duan, X.~Feng, M.~Gong, L.~Shou, B.~Qin, T.~Liu, D.~Jiang \emph{et~al.}, ``Codebert: A pre-trained model for programming and natural languages,'' in \emph{Findings of the Association for Computational Linguistics: EMNLP 2020}, 2020, pp. 1536--1547.

\bibitem{fraser2011evosuite}
G.~Fraser and A.~Arcuri, ``Evosuite: automatic test suite generation for object-oriented software,'' in \emph{Proceedings of the 19th ACM SIGSOFT symposium and the 13th European conference on Foundations of software engineering}, 2011, pp. 416--419.

\bibitem{pacheco2007randoop}
C.~Pacheco and M.~D. Ernst, ``Randoop: feedback-directed random testing for java,'' in \emph{Companion to the 22nd ACM SIGPLAN conference on Object-oriented programming systems and applications companion}, 2007, pp. 815--816.

\bibitem{aleti2017analysing}
A.~Aleti, I.~Moser, and L.~Grunske, ``Analysing the fitness landscape of search-based software testing problems,'' \emph{Automated Software Engineering}, vol.~24, pp. 603--621, 2017.

\bibitem{oliveira2018mapping}
C.~Oliveira, A.~Aleti, L.~Grunske, and K.~Smith-Miles, ``Mapping the effectiveness of automated test suite generation techniques,'' \emph{IEEE Transactions on Reliability}, vol.~67, no.~3, pp. 771--785, 2018.

\bibitem{panichella2015reformulating}
A.~Panichella, F.~M. Kifetew, and P.~Tonella, ``Reformulating branch coverage as a many-objective optimization problem,'' in \emph{2015 IEEE 8th international conference on software testing, verification and validation (ICST)}.\hskip 1em plus 0.5em minus 0.4em\relax IEEE, 2015, pp. 1--10.

\bibitem{panichella2017automated}
------, ``Automated test case generation as a many-objective optimisation problem with dynamic selection of the targets,'' \emph{IEEE Transactions on Software Engineering}, vol.~44, no.~2, pp. 122--158, 2017.

\bibitem{deepseek-coder}
D.~AI, ``Deepseek coder: Let the code write itself,'' \url{https://github.com/deepseek-ai/DeepSeek-Coder}, 2023.

\bibitem{roziere2023code}
B.~Rozi{\`e}re, J.~Gehring, F.~Gloeckle, S.~Sootla, I.~Gat, X.~E. Tan, Y.~Adi, J.~Liu, T.~Remez, J.~Rapin \emph{et~al.}, ``Code llama: Open foundation models for code,'' \emph{arXiv preprint arXiv:2308.12950}, 2023.

\bibitem{wang2021codet5}
Y.~Wang, W.~Wang, S.~Joty, and S.~C. Hoi, ``Codet5: Identifier-aware unified pre-trained encoder-decoder models for code understanding and generation,'' \emph{arXiv preprint arXiv:2109.00859}, 2021.

\bibitem{li2023starcoder}
R.~Li, L.~B. Allal, Y.~Zi, N.~Muennighoff, D.~Kocetkov, C.~Mou, M.~Marone, C.~Akiki, J.~Li, J.~Chim \emph{et~al.}, ``Starcoder: may the source be with you!'' \emph{arXiv preprint arXiv:2305.06161}, 2023.

\bibitem{JavaParser}
\BIBentryALTinterwordspacing
``Javaparser,'' 2024. [Online]. Available: \url{https://javaparser.org/}
\BIBentrySTDinterwordspacing

\bibitem{liu2023enhancing}
W.~Liu, S.~Cheng, D.~Zeng, and H.~Qu, ``Enhancing document-level event argument extraction with contextual clues and role relevance,'' \emph{arXiv preprint arXiv:2310.05991}, 2023.

\bibitem{liu2024beyond}
W.~Liu, L.~Zhou, D.~Zeng, Y.~Xiao, S.~Cheng, C.~Zhang, G.~Lee, M.~Zhang, and W.~Chen, ``Beyond single-event extraction: Towards efficient document-level multi-event argument extraction,'' \emph{arXiv preprint arXiv:2405.01884}, 2024.

\bibitem{miyato2016adversarial}
T.~Miyato, A.~M. Dai, and I.~Goodfellow, ``Adversarial training methods for semi-supervised text classification,'' \emph{arXiv preprint arXiv:1605.07725}, 2016.

\bibitem{wu2021r}
L.~Wu, J.~Li, Y.~Wang, Q.~Meng, T.~Qin, W.~Chen, M.~Zhang, T.-Y. Liu \emph{et~al.}, ``R-drop: Regularized dropout for neural networks,'' \emph{Advances in Neural Information Processing Systems}, vol.~34, pp. 10\,890--10\,905, 2021.

\bibitem{gao2021simcse}
T.~Gao, X.~Yao, and D.~Chen, ``Simcse: Simple contrastive learning of sentence embeddings,'' \emph{arXiv preprint arXiv:2104.08821}, 2021.

\bibitem{yin2024enhancing}
X.~Yin, C.~Ni, X.~Xu, X.~Li, and X.~Yang, ``Enhancing discriminative tasks by guiding the pre-trained language model with large language model's experience,'' \emph{arXiv preprint arXiv:2408.08553}, 2024.

\bibitem{bang2023multitask}
Y.~Bang, S.~Cahyawijaya, N.~Lee, W.~Dai, D.~Su, B.~Wilie, H.~Lovenia, Z.~Ji, T.~Yu, W.~Chung \emph{et~al.}, ``A multitask, multilingual, multimodal evaluation of chatgpt on reasoning, hallucination, and interactivity,'' \emph{arXiv preprint arXiv:2302.04023}, 2023.

\bibitem{yin2024rectifier}
X.~Yin, C.~Ni, T.~N. Nguyen, S.~Wang, and X.~Yang, ``Rectifier: Code translation with corrector via llms,'' \emph{arXiv preprint arXiv:2407.07472}, 2024.

\bibitem{ouyang2022training}
L.~Ouyang, J.~Wu, X.~Jiang, D.~Almeida, C.~Wainwright, P.~Mishkin, C.~Zhang, S.~Agarwal, K.~Slama, A.~Ray \emph{et~al.}, ``Training language models to follow instructions with human feedback,'' \emph{Advances in Neural Information Processing Systems}, vol.~35, pp. 27\,730--27\,744, 2022.

\bibitem{yin2024thinkrepair}
X.~Yin, C.~Ni, S.~Wang, Z.~Li, L.~Zeng, and X.~Yang, ``Thinkrepair: Self-directed automated program repair,'' in \emph{Proceedings of the 33rd ACM SIGSOFT International Symposium on Software Testing and Analysis}, 2024, pp. 1274--1286.

\bibitem{vaswani2017attention}
A.~Vaswani, N.~Shazeer, N.~Parmar, J.~Uszkoreit, L.~Jones, A.~N. Gomez, {\L}.~Kaiser, and I.~Polosukhin, ``Attention is all you need,'' \emph{Advances in neural information processing systems}, vol.~30, 2017.

\bibitem{zeng2022extensive}
Z.~Zeng, H.~Tan, H.~Zhang, J.~Li, Y.~Zhang, and L.~Zhang, ``An extensive study on pre-trained models for program understanding and generation,'' in \emph{Proceedings of the 31st ACM SIGSOFT international symposium on software testing and analysis}, 2022, pp. 39--51.

\bibitem{wan2022they}
Y.~Wan, W.~Zhao, H.~Zhang, Y.~Sui, G.~Xu, and H.~Jin, ``What do they capture? a structural analysis of pre-trained language models for source code,'' in \emph{Proceedings of the 44th International Conference on Software Engineering}, 2022, pp. 2377--2388.

\bibitem{liu2023towards}
Z.~Liu, K.~Liu, X.~Xia, and X.~Yang, ``Towards more realistic evaluation for neural test oracle generation,'' in \emph{Proceedings of the 32th International Symposium on Software Testing and Analysis}.\hskip 1em plus 0.5em minus 0.4em\relax ACM, 2023.

\bibitem{wang2023codet5+}
Y.~Wang, H.~Le, A.~D. Gotmare, N.~D. Bui, J.~Li, and S.~C. Hoi, ``Codet5+: Open code large language models for code understanding and generation,'' \emph{arXiv preprint arXiv:2305.07922}, 2023.

\bibitem{bessey2010few}
A.~Bessey, K.~Block, B.~Chelf, A.~Chou, B.~Fulton, S.~Hallem, C.~Henri-Gros, A.~Kamsky, S.~McPeak, and D.~Engler, ``A few billion lines of code later: using static analysis to find bugs in the real world,'' \emph{Communications of the ACM}, vol.~53, no.~2, pp. 66--75, 2010.

\bibitem{johnson2013don}
B.~Johnson, Y.~Song, E.~Murphy-Hill, and R.~Bowdidge, ``Why don't software developers use static analysis tools to find bugs?'' in \emph{2013 35th International Conference on Software Engineering (ICSE)}.\hskip 1em plus 0.5em minus 0.4em\relax IEEE, 2013, pp. 672--681.

\bibitem{pytorch}
\BIBentryALTinterwordspacing
A.~Paszke, S.~Gross, F.~Massa, A.~Lerer, J.~Bradbury, G.~Chanan, T.~Killeen, Z.~Lin, N.~Gimelshein, L.~Antiga, A.~Desmaison, A.~Kopf, E.~Yang, Z.~DeVito, M.~Raison, A.~Tejani, S.~Chilamkurthy, B.~Steiner, L.~Fang, J.~Bai, and S.~Chintala, ``Pytorch: An imperative style, high-performance deep learning library,'' in \emph{Advances in Neural Information Processing Systems 32}.\hskip 1em plus 0.5em minus 0.4em\relax Curran Associates, Inc., 2019, pp. 8024--8035. [Online]. Available: \url{http://papers.neurips.cc/paper/9015-pytorch-an-imperative-style-high-performance-deep-learning-library.pdf}
\BIBentrySTDinterwordspacing

\bibitem{huggingface}
\BIBentryALTinterwordspacing
``Hugging face,'' 2023. [Online]. Available: \url{https://huggingface.co}
\BIBentrySTDinterwordspacing

\bibitem{ni2022defect}
C.~Ni, K.~Yang, X.~Xia, D.~Lo, X.~Chen, and X.~Yang, ``Defect identification, categorization, and repair: Better together,'' \emph{arXiv preprint arXiv:2204.04856}, 2022.

\bibitem{menzies2006data}
T.~Menzies, J.~Greenwald, and A.~Frank, ``Data mining static code attributes to learn defect predictors,'' \emph{IEEE transactions on software engineering}, vol.~33, no.~1, pp. 2--13, 2006.

\bibitem{zimmermann2007predicting}
T.~Zimmermann, R.~Premraj, and A.~Zeller, ``Predicting defects for eclipse,'' in \emph{Third International Workshop on Predictor Models in Software Engineering (PROMISE'07: ICSE Workshops 2007)}.\hskip 1em plus 0.5em minus 0.4em\relax IEEE, 2007, pp. 9--9.

\bibitem{nagappan2008influence}
N.~Nagappan, B.~Murphy, and V.~Basili, ``The influence of organizational structure on software quality: an empirical case study,'' in \emph{Proceedings of the 30th international conference on Software engineering}, 2008, pp. 521--530.

\bibitem{nagappan2005use}
N.~Nagappan and T.~Ball, ``Use of relative code churn measures to predict system defect density,'' in \emph{Proceedings of the 27th international conference on Software engineering}.\hskip 1em plus 0.5em minus 0.4em\relax ACM, 2005, pp. 284--292.

\bibitem{graves2000predicting}
T.~L. Graves, A.~F. Karr, J.~S. Marron, and H.~Siy, ``Predicting fault incidence using software change history,'' \emph{IEEE Transactions on software engineering}, vol.~26, no.~7, pp. 653--661, 2000.

\bibitem{kim2007predicting}
S.~Kim, T.~Zimmermann, E.~J. Whitehead~Jr, and A.~Zeller, ``Predicting faults from cached history,'' in \emph{29th International Conference on Software Engineering (ICSE'07)}.\hskip 1em plus 0.5em minus 0.4em\relax IEEE, 2007, pp. 489--498.

\bibitem{ni2022best}
C.~Ni, W.~Wang, K.~Yang, X.~Xia, K.~Liu, and D.~Lo, ``The best of both worlds: integrating semantic features with expert features for defect prediction and localization,'' in \emph{Proceedings of the 30th ACM Joint European Software Engineering Conference and Symposium on the Foundations of Software Engineering}, 2022, pp. 672--683.

\bibitem{zhao2023systematic}
Y.~Zhao, K.~Damevski, and H.~Chen, ``A systematic survey of just-in-time software defect prediction,'' \emph{ACM Computing Surveys}, vol.~55, no.~10, pp. 1--35, 2023.

\bibitem{misirli2016studying}
A.~T. Misirli, E.~Shihab, and Y.~Kamei, ``Studying high impact fix-inducing changes,'' \emph{Empirical Software Engineering}, vol.~21, pp. 605--641, 2016.

\bibitem{pornprasit2021jitline}
C.~Pornprasit and C.~K. Tantithamthavorn, ``Jitline: A simpler, better, faster, finer-grained just-in-time defect prediction,'' in \emph{2021 IEEE/ACM 18th International Conference on Mining Software Repositories (MSR)}.\hskip 1em plus 0.5em minus 0.4em\relax IEEE, 2021, pp. 369--379.

\bibitem{pornprasit2021pyexplainer}
C.~Pornprasit, C.~Tantithamthavorn, J.~Jiarpakdee, M.~Fu, and P.~Thongtanunam, ``Pyexplainer: Explaining the predictions of just-in-time defect models,'' in \emph{2021 36th IEEE/ACM International Conference on Automated Software Engineering (ASE)}.\hskip 1em plus 0.5em minus 0.4em\relax IEEE, 2021, pp. 407--418.

\end{thebibliography}
